\documentclass[aps, pre,amsmath,amssymb,twocolumn,floatfix]{revtex4}
\usepackage{graphicx}
\usepackage{epsfig}
\input epsf

\newcommand {\dr}{{\mathrm d}\mathbf{r}}

\newcommand {\rr}{\mathbf{r}}

\newcommand {\kk}{\mathbf{k}}
\newcommand {\etal}{\begin{itshape}et al\end{itshape}.}

\newcommand {\ie}{$\epsilon^{-1}$}

\begin{document}

\title{\bf Phase behavior of a fluid with competing attractive and repulsive interactions}

\author{Andrew J.~Archer$^{1,2}$}
\author{Nigel B.~Wilding$^2$}
\affiliation{1.~Department of Mathematical Sciences, Loughborough University, Loughborough LE11 3TU, United Kingdom\\
2.~Department of Physics, University of Bath, Bath BA2 7AY, United Kingdom}

\begin{abstract} 

Fluids in which the interparticle potential has a hard core, is attractive at
moderate separations, and repulsive at greater separations are known
to exhibit novel phase behavior, including stable inhomogeneous phases.
Here we report a joint simulation and theoretical study of such a
fluid, focusing on the relationship between the liquid-vapor transition
line and any new phases. The phase diagram is studied as a function of
the amplitude of the attraction for a certain fixed amplitude of the
long ranged repulsion. We find that the effect of the repulsion is to
substitute the liquid-vapor critical point and a portion of the
associated liquid-vapor transition line, by two first order
transitions. One of these transitions separates the vapor from a fluid
of spherical liquidlike clusters; the other separates the liquid from a
fluid of spherical voids. At low temperature, the two
transition lines intersect one another and a vapor-liquid transition
line at a triple point. While most integral equation theories are
unable to describe the new phase transitions, the Percus Yevick
approximation does succeed in capturing the vapor-cluster transition,
as well as aspects of the structure of the cluster fluid, in reasonable
agreement with the simulation results.

\end{abstract} 
\maketitle 
\epsfclipon  


\section{Introduction}

Until quite recently, it was commonly supposed that the gamut of
equilibrium fluid phase behavior exhibited by a single component system
of particles interacting via an isotropic pair potential, does not
extend beyond a liquid phase and a vapor phase. While this certainly
appears to be true for prototype models such as the Lennard-Jones
fluid, it is now recognized that considerably richer phase behavior can
occur \cite{MalescioJPCM2007}. For instance, systems whose particles
have a repulsive core that is sufficiently `soft', may exhibit one or
more stable liquid-liquid phase transitions over and above the
liquid-vapor transition \cite{Jagla99,WildingMagee02,Gibson06}.
Furthermore, and notwithstanding their isotropy, such soft
potentials can give rise to inhomogeneous fluid phases composed of clusters,
or patterned morphologies such as stripes and lamellae
\cite{MladeketalPRL2006, MalescioandPellicaneNatureMatt2003,
SearFrenkelPRL2003,CampPRE2003}.

Another class of isotropic pair potential exhibiting rich
phase behavior is the ``mermaid'' potential--so called because it is
attractive at moderate distances and repulsive at large distances
(i.e.\ it has an attractive `head' and a repulsive `tail')
\cite{PinietalJPCM2006}. Effective potentials of the mermaid form
can be found in colloid-polymer mixtures
\cite{StradneretalNature2004, SedgwicketalJPCM2004,
Bartlett,Bartlett2}: The long ranged repulsion arises from the weakly
screened charge carried by the colloids, while the attraction at
intermediate distances stems from the depletion forces associated with
the non-adsorbing polymers \cite{BarratHansen}. The competition between
attraction and repulsion on distinct length scales is responsible for
novel behavior such as the appearance of equilibrium cluster phases
and non-equilibrium gel states \cite{SedgwicketalJPCM2004,Bartlett}.
Mermaid potentials are further appropriate for the description of protein
solutions \cite{StradneretalNature2004,BordietalBiophysJ2006}, 
star-polymer systems \cite{LoVersoetalProgrColloidPolymSci2006}, the
effective interactions between solute particles in a subcritical liquid solvent
\cite{ChakrabartietalJPCM2006}, and colloidal monolayers
\cite{GhezziEarnshawJPCM1997, SearetalPRE1999}. In the latter (two
dimensional) systems, the short range attraction arises from the van
der Waals and capillary forces, while the longer range repulsion is
thought to be due to dipole-dipole interactions
\cite{GhezziEarnshawJPCM1997,SearetalPRE1999}. Such two dimensional
systems have also been observed to form cluster and stripe morphologies
\cite{GhezziEarnshawJPCM1997, SearetalPRE1999,
ImperioReattoJPCM2004,ImperioReattoJCP2006}. 

Beyond their relevance to the real systems listed above, models
whose particles interact via a mermaid potential are of considerable
interest from the fundamental perspective of statistical mechanics.
Specifically, they elicit basic questions such as: What is the scope
and character of their phase behaviour, and how can this be described
theoretically? In what follows, we briefly review the progress to date
in addressing these issues.

A number of theoretical and simulation studies have considered aspects
of the phase behaviour of a variety of fluids interacting via mermaid
potentials \cite{AndelamnetalJCP1987, KendricketalEPL1988, 
SeulAndelmanScience1995, NussinovetalPRL1999, MuratovPRE2002,
TarziaConiglioPRE2007, SearGelbartJCP1999, GroenewoldKegelJPCB2001,
GroenewoldKegelJPCM2004, PinietalCPL2000, ARCHER19, WuetalPRE2004,
LiuetalJCP2005, BroccioetalJCP2006, SciortinoetalPRL2004,
MossaetalLangmuir2004, SciortinoetalJPCB2005, CandiaetalPRE2006,
CHARBONNEAU07}. In pioneering early work \cite{AndelamnetalJCP1987,
KendricketalEPL1988, SeulAndelmanScience1995}, a mean-field (Landau)
theory approach was developed for systems with competing interactions.
This predicted that when the amplitude of the long ranged repulsion is
sufficiently large (relative to that of the attraction), modulated
phases appear in the region of the phase diagram where, if one were only
to allow for the occurrence of homogeneous phases, the theory would predict the
liquid-vapor critical point to be located. More recent field-theoretic
studies \cite{NussinovetalPRL1999, MuratovPRE2002,
TarziaConiglioPRE2007} have arrived at similar conclusions. 

Adopting a phenomenological approach, Groenewold and Kegel
\cite{GroenewoldKegelJPCB2001, GroenewoldKegelJPCM2004} developed a
model to explain how competition between short ranged attraction and a
longer ranged repulsion in colloidal systems could promote cluster
formation. They concluded that depending on the relative strengths and
ranges of the competing attractive and repulsive contributions to the
pair potential, large clusters (up to several thousands of particles)
would be stable. Signatures of such clustering were observed by Sear and
Gelbart within a mean-field (random phase approximation (RPA)) liquid
state theory \cite{SearGelbartJCP1999} study of a model in which the
attractive and repulsive contributions to the potential are both
rather long ranged (justifying the mean-field approximation). They
showed that the propensity to clustering is manifest by a peak in the
static structure factor, $S(k)$, at a small but non-zero wavevector
$k_c$. In fact, the mean-field theory predicts a line in the phase
diagram at which $S(k)$ {\em diverges} at $k=k_c \neq 0$, and it was
inferred that this line indicates microphase separation to a modulated
phase(s). Such a line has previously been dubbed the ``$\lambda$-line''
\cite{ARCHER19} in terminology borrowed from other contexts
\cite{ARCHER04,STELL95,CIACH03}. 

Within the framework of integral equation theory, Chen and coworkers
\cite{WuetalPRE2004, LiuetalJCP2005, BroccioetalJCP2006} solved the
Ornstein-Zernike (OZ) equation for the structure of a fluid of particles
interacting via a mermaid potential in which the attractive and
repulsive contributions are both assigned the Yukawa form (a
``double-Yukawa'' potential) \cite{HM}. Two closures were examined: the
mean spherical approximation (MSA) and the hypernetted-chain (HNC)
approximation. Focusing on the portion of the phase diagram where the
peak in $S(k)$ at $k=k_c$ is developing, the authors found that both
theories provide a good description of the fluid structure, as gauged
by comparison with Monte Carlo (MC) simulations. As will be demonstrated
in the present work, however, neither the MSA nor the HNC is reliable for
describing the fluid structure in the vicinity of transitions to
inhomogeneous phases. 

A generally more accurate integral equation theory, the self consistent
Ornstein-Zernike approximation (SCOZA), has been applied in conjunction
with hierarchical reference theory (HRT) calculations,  by Pini \etal\
\cite{PinietalCPL2000, PinietalJPCM2006} to a double-Yukawa mermaid
potential. The authors investigated the influence on the structure and
phase behaviour of the fluid as the amplitude of the repulsive
contribution was increased from zero. Doing so was found to depress the
vapor-liquid critical temperature and led to the appearance of an
anomalously large region around the liquid-vapor critical point in
which the fluid compressibility is very high \cite{PinietalCPL2000,
PinietalJPCM2006}. Unfortunately, Pini \etal\ were unable to obtain
solutions either from the SCOZA or the HRT in the regime where
inhomogeneous phases might be expected to occur \cite{PinietalCPL2000,
PinietalJPCM2006, ARCHER19}. The SCOZA results were subsequently
compared with a RPA density functional theory (DFT) by Archer {\em et
al} \cite{ARCHER19}. In common with earlier mean field approaches
\cite{SearGelbartJCP1999}, the RPA DFT predicts a $\lambda$-line to
occur in the phase diagram when the amplitude of repulsive part of the
potential exceeds a threshold value \cite{ARCHER19}. In parts of the
phase diagram away from the $\lambda$-line, reasonable agreement with
SCOZA was found for the fluid structure and thermodynamics.

Sciortino and co-workers \cite{SciortinoetalPRL2004,
MossaetalLangmuir2004, SciortinoetalJPCB2005} have employed a variety
of theoretical and simulation techniques to investigate the properties
of mermaid systems. They focused attention on potentials having a
particularly short ranged attractive part (i.e.\ a small fraction of
the particle diameter), chosen to mimic the effective pair potential in
certain colloidal systems. They found -- as has been shown
experimentally \cite{Bartlett} -- that the system exhibits both a fluid
cluster phase and a gel phase comprising interconnected chains composed
of face-sharing tetrahedral clusters. Indeed, as has been recently
emphasized \cite{CHARBONNEAU07,TarziaConiglioPRE2007}, it is a general
feature of models having particularly short ranged attraction, that the
microphase separation can occur in a regime where the system dynamics
are very slow and a transition to a non-ergodic state is possible.
Another mermaid potential having a similarly short ranged attraction
was studied using simulation by de Candia \etal\, who observed columnar
and lamellar phases \cite{CandiaetalPRE2006}. It has even been
suggested that for such very short ranged attractive potentials, the
presence of a long ranged repulsion might not be a prerequisite for
cluster formation \cite{LU06}.

Notwithstanding the extensive body of impressive results on a variety
of model mermaid systems, displaying an intriguing wealth of
inhomogeneous phases, central questions remain unanswered.
Specifically, the detailed relationship between the liquid-vapor
transition and the inhomogeneous phases still seems obscure (recall
that mean-field theories predict a $\lambda$-line enclosing a region of
the phase diagram in which a `naive' application of the theory would
predict the liquid vapor critical point to be \cite{ARCHER19}.)
Furthermore, if the vapor-liquid critical point {\em is} lost when long
ranged repulsion is introduced, what happens to the remainder of the
liquid-vapor transition line?  In the present work, we attempt to
answer these questions by deploying simulation and theory to
investigate the structure and phase behavior of a model system whose
particles interact via a double-Yukawa mermaid potential. 

Our principal findings are as follows. Our Monte Carlo simulations show
that for a certain (moderately large) strength of the repulsive
contribution to the pair potential, the liquid-vapor critical point is
absent. In its stead we find two lines of first order phase
transitions, each of which separates a homogeneous phase from an
inhomogeneous (cluster) phase. One of these two transition lines is
located at low particle number density, and separates the vapor from a
fluid of spherical liquidlike clusters; it appears to terminate at a
critical point at high temperatures. The other line -- located at high
density -- separates a phase of spherical voids from the
homogeneous liquid; it too appears to terminate at a critical point. At
low temperature, the two transition lines intersect one another and a
vapor-liquid transition line at a triple point. 

We complement our simulation studies with an investigation of the
utility of a number of standard liquid state theories for describing
the phase behaviour of our model. We first apply the DFT of
Ref.~\cite{ARCHER19} to trace the locus of the $\lambda$-line, noting
that if one interprets this line as representing the transition to
periodically modulated phases, then its topology is incompatible with
the phase diagram emerging from the simulations. Turning our attention
to integral equation theories, we find that the HNC has a no-solution
region in the portion of the phase diagram where the transitions to
inhomogeneous phases occur, although it does yield a portion of the
vapor-liquid transition. Use of a simple modified HNC (MHNC)
approximation similarly fails to provide a solution in the region of
interest. We further find that (for separate reasons) the MSA is of
little use in this region of the phase diagram. Interestingly, however,
the Percus Yevick (PY) approximation {\em is} able to describe the
vapor-cluster phase transition as well as key aspects of the structure
of the cluster fluid, in reasonable agreement with the MC results.

Our paper is organized as follows. We introduce our model mermaid
potential in Sec.~\ref{sec:model}. The MC simulation
methodology used to study the model, and our findings concerning the
phase behaviour and the character of the inhomogeneous phases are
described in Sec.~\ref{sec:sim}. An investigation of the utility of
mean field DFT and integral equations for describing the phase behavior
is detailed in Sec.~\ref{sec:mf}. Finally, a discussion of the
implications of our findings and the outlook for future work features
in Sec.~\ref{sec:disc}.

\section{Model}

The model that we have elected to study comprises an isotropic two-body
interparticle potential of the hard-core plus double-Yukawa form:

\label{sec:model}

\begin{equation}
\beta v(r) = 
\begin{cases}
\infty \hspace{45mm} r \leq \sigma \\
-\epsilon \sigma \exp(-z_1(r/\sigma-1))/r\\
 \hspace{3mm}+ A \sigma \exp(-z_2(r/\sigma-1))/r \hspace{5mm} r > \sigma.
\end{cases}
\label{eq:pair_pot}
\end{equation}
Here $\beta=1/k_BT$ is the inverse temperature, which we set equal to
unity, while $\sigma$ is the particle diameter. The first Yukawa term
represents an interparticle attraction whose strength is controlled by
a parameter $\epsilon>0$, while the second term, with strength
parameter $A>0$, represents a repulsion.  We shall focus on the regime in which the range of
the repulsion exceeds that of the attraction i.e.\ $z_1>z_2$, leading
to a repulsive tail to the potential. 

The specific choice of $z_1$ and $z_2$ requires a balance to be struck
between a number of competing desiderata. Firstly, the attractive part
should not be so short-ranged that the equilibrium liquid-vapor phase
behavior is wholly preempted by freezing \cite{PinietalJPCM2006,
HAGEN94} and/or dynamic arrest (gelation) \cite{CHARBONNEAU07},
phenomena which -- while undoubtedly of interest in their own right -- are
not the specific concern of this work. Secondly, the range of the
repulsive part must exceed that of the attractive contribution, and be
sufficiently large to enable meaningful comparisons with mean field
theories, the accuracy of which increases with the potential range.
Finally, from a purely practical standpoint, the range should not be so
great that very large system sizes are necessary in order to obviate
cutoff artifacts when the potential is truncated in the interests of
computational efficiency. A satisfactory compromise in these respects
was found by assigning $z_1=2, z_2=1$, and truncating the potential at
$r=r_c=7.0\sigma$. For reasons relating to the occurrence of large
length-scale inhomogeneous phases, no mean field correction was applied
for the effects of the potential truncation.

As described in the Introduction, physically one can regard the
potential of Eq.~(\ref{eq:pair_pot}) as capturing the combined overall
effect of a screened coulombic repulsion between charged colloidal
particles, and a shorter-ranged attraction engendered by a depletion
agent such as polymer chains. In this spirit, it is natural to
regard $\epsilon$ as a measure of the polymer fugacity, $A$ as a
measure of the colloidal charge, and treat both parameters
independently. In the present work, however, we shall primarily be
concerned with the phase behavior as a function of \ie\ at fixed
$A$. For the latter, we have assigned the value $A=0.55$. This choice
was motivated by a preliminary study of the model within DFT
\cite{ARCHER19}, which suggested that (given the choice
$z_1=2, z_2=1$) this strength of repulsion is sufficient
to engender a $\lambda$-line encompassing the critical region of the
liquid-vapor line. One might therefore hope that qualitative
alterations to the standard scenario of liquid-vapor phase behavior would ensue.

\begin{figure}[h]
\includegraphics[width=0.98\columnwidth,clip=true]{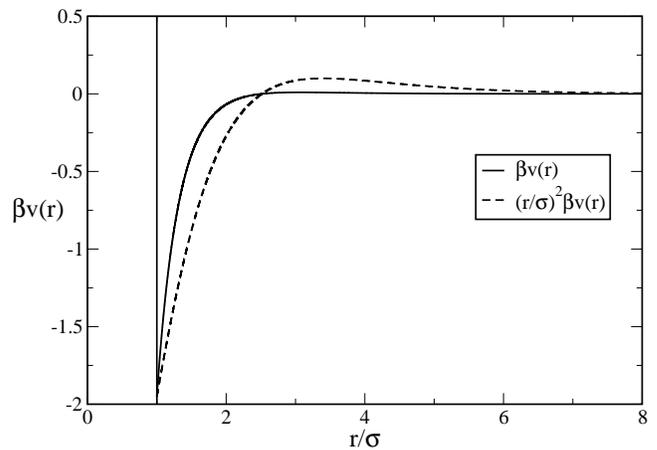}
\caption{The form of the potential $\beta v(r)$
(Eq.~\protect\ref{eq:pair_pot}) for parameters $\epsilon^{-1}=0.4,
A=0.55, z_1=2, z_2=1$. Also shown for comparison is the form of $(r/\sigma)^2\beta v(r)$.}
\label{fig:pot}
\end{figure}

The form of the potential $\beta v(r)$ is shown in Fig.~\ref{fig:pot}
for parameter values $\epsilon^{-1}=0.4, A=0.55, z_1=2, z_2=1$. Also
shown is the form of $(r/\sigma)^2\beta v(r)$, which provides a useful
indication of which cutoff values, $r_c$, are likely to result in
significant corrections to the internal energy compared to the full
potential. The figure confirms that for our choice of the cutoff,
$r_c=7.0\sigma$, only small corrections to the limiting behaviour are
to be expected. It should be noted, however, that because this cutoff
value is much larger than the values typically used in the simulation
of simple fluids such as Lennard-Jonesium, any simulation study is
expected to entail a considerable computational investment.

\section{Simulation studies}
\label{sec:sim}
\subsection{Techniques}

We have studied the phase behaviour of the model of
Sec.~\ref{sec:model}  using a grand canonical MC simulation algorithm
\cite{Frenkelsmit2002}. Where possible, accurate location of points of
first order phase coexistence was facilitated by the use of the
multicanonical preweighting technique \cite{berg1992}, aided by
multihistogram reweighting \cite{Ferrenberg1989} according to the
procedure described in ref.~\cite{Wilding2001}. Most of the results we
shall present were obtained for a system of linear size $L=21\sigma$,
although some data was also collected for $L=28\sigma$ in order to
gauge the scale of finite-size effects. Periodic boundaries were employed
throughout.

As shall be described below, techniques were implemented to determine
the distribution of sizes of particle clusters in selected regions of
the phase diagram. A cluster comprises a subset of particles that are
interlinked via pathways of interparticle `bonds'. However, in contrast
to lattice models, the definition of a bond in a system with continuous
translational symmetry is somewhat ambiguous. We adopt a criterion
which derives from that used for cluster identification in spin models.
Specifically, we determine the interaction energy $v$ between each pair
of particles and assign a bond with probability $p_{\rm
bond}=1-\exp(\beta v)$. Clusters of bonded particles are then
identified using the efficient enumeration algorithm of Hoshen and Kopelman
\cite{Hoshen76}. 

\subsection{Phase diagram}
\label{sec:pd}

Prior to commencing exploration of the phase diagram, an initial
estimate was required for the range of values of the attractive
strength for which non-homogeneous states might be expected to occur.
Guided by the DFT calculations reported in Sec.~\ref{sec:mf} (which predict that
for $A=0.55$, a $\lambda$-line appears for $0.54 > \epsilon^{-1} >
0.48$), we selected \ie$=0.5$ as a suitable candidate. Subsequently it
was found that smaller values of \ie\ were necessary to generate
inhomogeneous phases.

\begin{figure}[h]
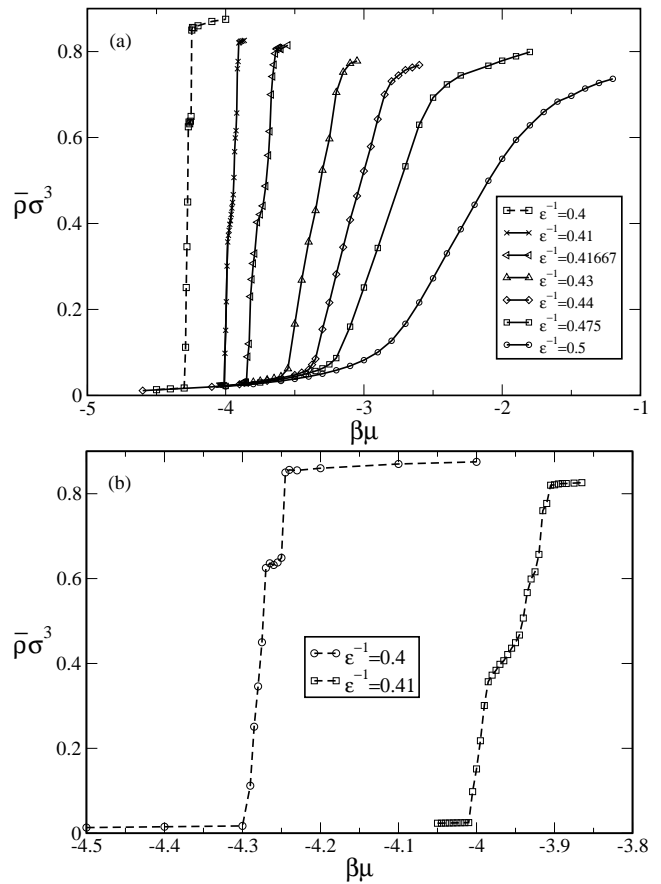

\includegraphics[width=0.98\columnwidth,clip=true]{fig2a.eps}
\includegraphics[width=0.98\columnwidth,clip=true]{fig2b.eps}
\caption{{\bf (a)} Measured forms of $\bar\rho(\mu)$ for $z_1=2$, $z_2=1$, $A=0.55$
and various \ie. {\bf (b)} An enlargement showing the data for \ie$=0.41$ and \ie$=0.4$
as described in the text.}
\label{fig:isotherms}
\end{figure}

For each value of \ie\ studied, the dependence of the average of the
fluctuating particle number density, $\bar\rho$, on the applied
chemical potential $\mu$ was measured. The resulting forms of
$\bar\rho(\mu)$ are shown in Fig.~\ref{fig:isotherms}(a). From this
figure one observes that for the initial value \ie$=0.5$,
$\bar\rho(\mu)$ is essentially smooth, having a form reminiscent of a
one-phase (super-critical) fluid. However, as \ie\ is reduced from this
value, the gradient of the curves for moderate densities gradually
increase in magnitude, indicating an increase in the fluid
compressibility. Concomitantly, kinks in the curves start to develop at
low and high density, which sharpen into discontinuities as \ie\ is
further decreased.  For \ie $\lesssim 0.42$, these low and high density
kinks are supplemented by additional ones at intermediate densities, as is
apparent from the close-up of the data for \ie$=0.41$ and \ie$=0.4$
shown in Fig.~\ref{fig:isotherms}(b).

We address first the phenomena underlying the appearance of the
discontinuities in $\bar\rho(\mu)$ at low and high density. These arise
from first order phase transitions. In the low density case, the
transition is between a vapor and an inhomogeneous phase composed of
large spherical liquidlike clusters; in the high density case it is
from a liquid to an inhomogeneous liquid phase containing large
spherical vaporlike voids (i.e.\  bubbles).  Discussion of the
character of these inhomogeneous structures is deferred until
Sec.~\ref{sec:character}. 

\begin{figure}[h]
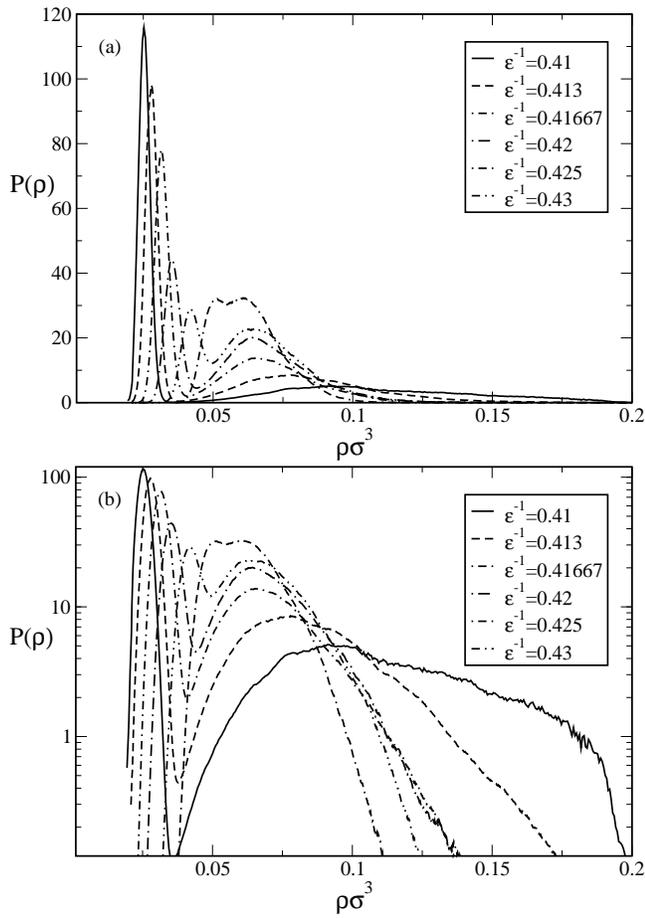

\includegraphics[width=0.98\columnwidth,clip=true]{fig3a.eps}
\includegraphics[width=0.98\columnwidth,clip=true]{fig3b.eps}
\caption{{\bf (a)} The distribution of the fluctuating number density for state
points corresponding to the low density vapor-spherical cluster
transition. {\bf (b)} The same data plotted on a log scale.}
\label{fig:lowertrans}
\end{figure}

Evidence for the existence of these phase transitions comes from the measured
forms of the distribution of the fluctuating instantaneous number
density $p(\rho)$ at the model parameters for which the kinks occur.
For the low density transition, these distributions are shown in
Fig.~\ref{fig:lowertrans}(a). One observes a two-peaked structure; the
narrow peak at low densities corresponds to the vapor phase, while a
much broader higher density peak corresponds to the cluster phase. At
small values of \ie\ the two peaks are widely separated, the trough
between them is deep (cf. the log scale of
fig.~\ref{fig:lowertrans}(b)), and the cluster peak is very broad (this
latter feature reflects the large compressibility associated with the
steepness of $\bar\rho(\mu)$ at these values of \ie\ --recall
Fig.~\ref{fig:isotherms}(a)). As \ie\ is increased, the two peaks
approach one another and eventually merge into a single peak; this occurs at
a value of \ie\ consistent with that at which the sharp low density kink
in $\bar\rho(\mu)$ becomes smoothed out.

\begin{figure}[h]
\includegraphics[width=0.98\columnwidth,clip=true]{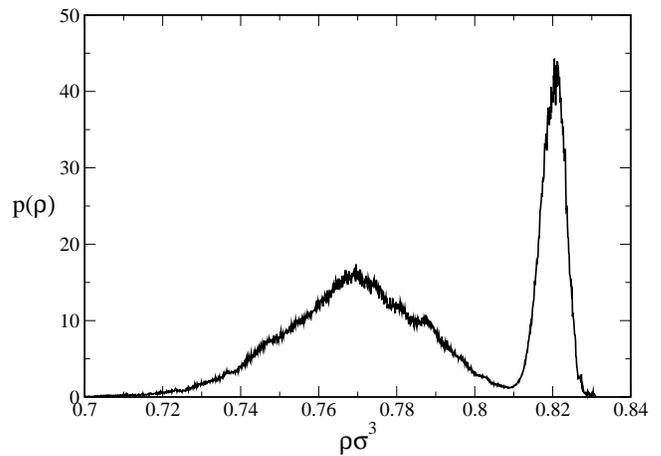}
\caption{The distribution of the fluctuating number density for the state
point $\epsilon^{-1}=0.41, \beta\mu=-3.91$, lying on the high density spherical
bubble-liquid transition.}
\label{fig:uppertrans}
\end{figure}

The computational cost of performing long simulations in the region of the
high density transition, verges on the prohibitive. Consequently, we have been
able to obtain the coexistence form of $p(\rho)$ for only a single value of
\ie, namely $\epsilon^{-1}=0.41$. This distribution is shown in
Fig.~\ref{fig:uppertrans} and exhibits a narrow liquidlike peak at high
density together with a much broader peak at lower densities which corresponds to a liquid
containing a large void, as will be discussed in
Sec.~\ref{sec:character}. Again the parameters at which we find the
double-peaked structure match those of the corresponding kink in
$\bar\rho(\mu)$, which can therefore be taken as an alternative signature
of the transition.

\begin{figure}
\includegraphics[width=0.98\columnwidth,clip=true]{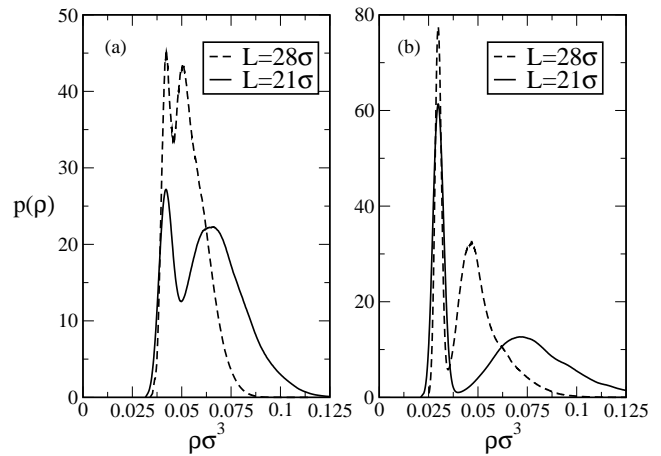}
\caption{Near-coexistence density distributions for {\bf
(a)} \ie$=0.415$ and {\bf (b)} \ie$=0.425$. In each case data is shown
for two system sizes.}
\label{fig:finitesize}
\end{figure}

Unfortunately, it proved impossible to accurately quantify the number
and energy densities of the cluster phases that coexist with the
respective vapor and liquid phases. The problems are traceable to the
large length scale associated with the typical cluster size, as 
illustrated in Fig.~\ref{fig:finitesize}(a), which shows the measured
density distribution $p(\rho)$ for two system sizes at a
near-coexistence point located well inside the two-phase vapor-cluster
region. From the figure, it is evident that while the position of the
vapor phase peak is essentially system size-independent, the density of
the cluster peak shifts strongly to lower density as $L$ is increased.
Thus the system sizes attainable in this work fail to provide access to the
thermodynamic limit for this phase. Analogous effects are found for the
energy distribution (data not shown).

\begin{figure}
\includegraphics[width=0.98\columnwidth,clip=true]{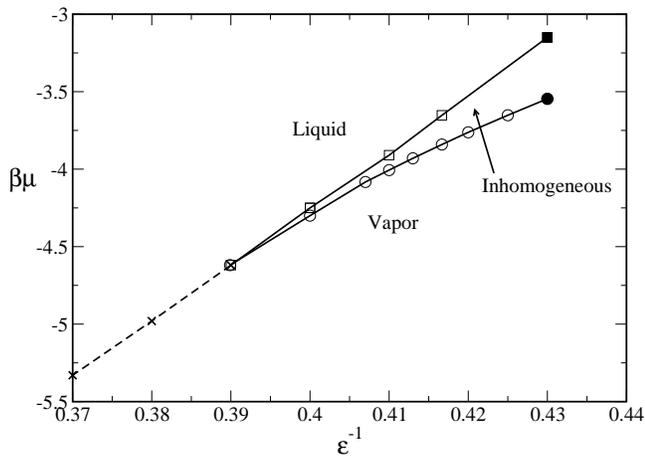}
\caption{Estimates of the phase diagram, as described in the text.
Circles are the vapor-spherical cluster transition, squares are the
spherical bubble-liquid transition. Crosses lie on the liquid-vapor
coexistence line, lines merely guide the eye. Filled symbols locate the putative critical
points. Uncertainties do not exceed the symbol sizes.}
\label{fig:coexcurve}
\end{figure}

The low and high density first order transition lines in $\epsilon-\mu$
space are presented in Fig.~\ref{fig:coexcurve}.  In this figure, the
estimates of points on the low density transition line derive from the
data of Fig.~\ref{fig:lowertrans} by locating the values of $\mu$ for
which $p(\rho)$ exhibits two peaks of approximately equal weight. Those
for the high density transition derive both from the data of
Fig.~\ref{fig:uppertrans} for the case \ie$=0.41$, and otherwise from
the position of the high density kink in the $\bar\rho(\mu)$ curves.
The loci of the two transition lines delimits a region within which
inhomogeneous structures occur. For small \ie $\simeq 0.39$, this
region tapers down to a point, beyond which we were unable to stabilize
inhomogeneous phases, instead finding a single transition from a vapor
to a liquid. This suggests that a triple point occurs at  \ie
$\simeq0.39$ below which standard vapor-liquid coexistence occurs. We
have estimated the locus of a portion of the vapor-liquid coexistence
from the center of the hysteresis loop in $\epsilon-\mu$ formed by
traversing the transition from vapor to liquid and back again, and
these estimates are also marked on Fig.~\ref{fig:coexcurve}. Owing to
the great number of particles in the liquid phase, the considerable potential
cutoff distance, and the substantial density difference between the liquid
and vapor, it was not possible to link the phase spaces of coexisting
vapor and liquid directly using biased sampling techniques, as was done
for the low density (vapor-cluster) transition.

As regards the nature of the transitions for large \ie, one observes
from Fig.~\ref{fig:lowertrans} that for the vapor-cluster transition,
the two peaks in $p(\rho)$ coalesce at a value of \ie\ close to
that at which the low density kink in $\bar\rho(\mu)$ disappears. It seems
reasonable to assume that a critical point occurs in this region and
that, by extension, the high density transition similarly end at a critical
point near to where the high density kink in $\bar\rho(\mu)$ vanishes.
However, accurately pinning down the respective critical point
parameters, proved problematic. The source of the difficulty is, again,
the large cluster length scale. Specifically, on increasing \ie, the
two peaks merge before any system size dependence of the
vapor peak becomes apparent. This implies that (in contrast to the case
for e.g.\ a liquid-vapor transition in a simple fluid
\cite{wilding1995a}) the double peaked structure of $p(\rho)$ is lost
{\em before} the thermal correlation length becomes comparable with the
system size. Accordingly, the value of \ie\ at which the peaks merge in
a finite-sized system constitutes an {\em underestimate} of the true
critical point value of \ie\ and can thus, strictly speaking, only
provide a lower bound on the critical point value of \ie
\cite{Archerfootnote2}. The existence of an additional large length
scale for one of the phases severely complicates the implementation of
standard finite-size methods for locating criticality, which are based
on the assumption of a single dominant length scale in the
near-critical region, namely the thermal correlation length. 
Nevertheless, for \ie $> 0.43$ the low and high density kinks in
$\bar\rho(\mu)$ have disappeared, the curves are smooth, and we thus
tentatively (and conservatively) assign $0.425 \le \epsilon_c^{-1}\le 0.440$ for
both the low density and high density transitions.

\subsection{Nature of the coexisting phases at the low and high
density transitions}
\label{sec:character}

\begin{figure}[h]
\includegraphics[width=0.98\columnwidth,clip=true]{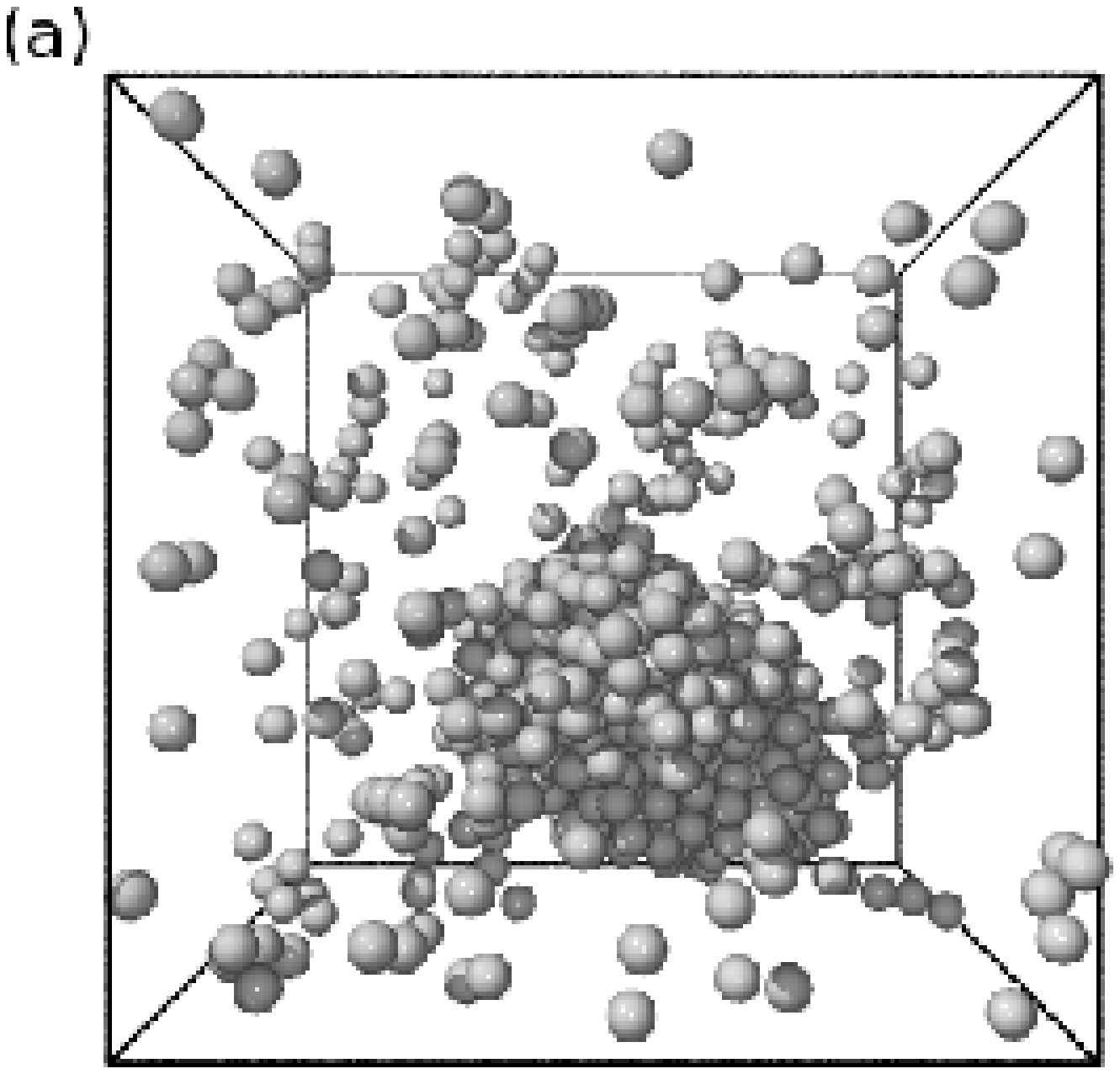}
\includegraphics[width=0.98\columnwidth,clip=true]{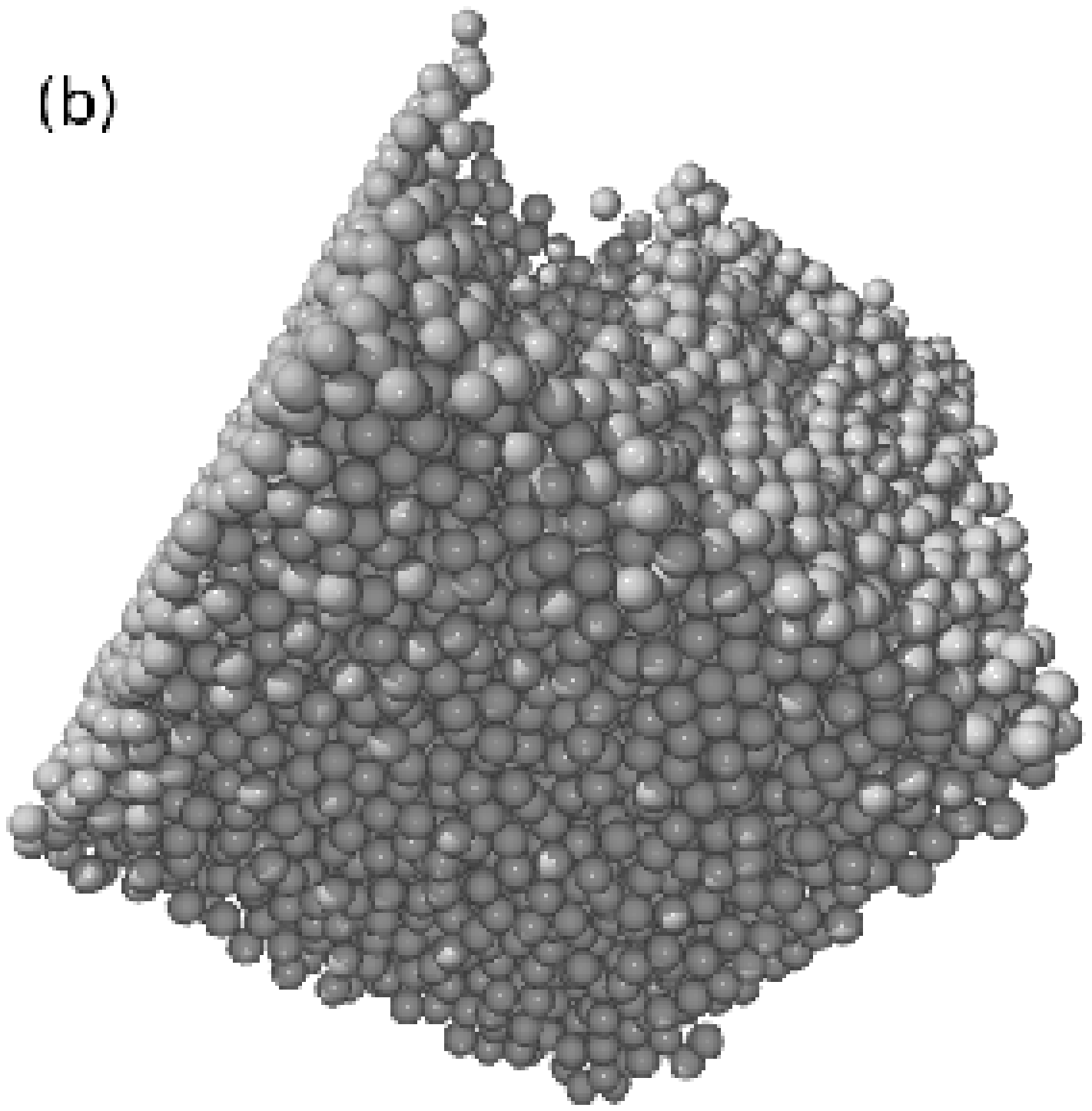}
\caption{{\bf (a)} A typical configurations of the spherical cluster phase
which coexists with vapor (not shown) at the low density transition. {\bf (b)} A typical configuration of the spherical
vapor bubble phase that coexists with homogeneous liquid (not shown) at the high density transition.}
\label{fig:coexist}
\end{figure}

In Sec.~\ref{sec:pd} we mapped the boundary between the homogeneous and
inhomogeneous phases. Here we ponder the character of the
inhomogeneous phases in more detail. Fig.~\ref{fig:coexist}(a) and
(b) show, for \ie$=0.41$, typical configuration snapshots of the
inhomogeneous phase which coexists with homogeneous vapor (liquid) at the low
(high) density transitions respectively. For the low density
transition, a vapor coexists with a phase containing an approximately
spherical cluster, whose local density is liquidlike. For the high
density transition, a dense liquid coexists with a phase containing a
large spherical bubble of vapor. 

\begin{figure}[h]
\includegraphics[width=0.98\columnwidth,clip=true]{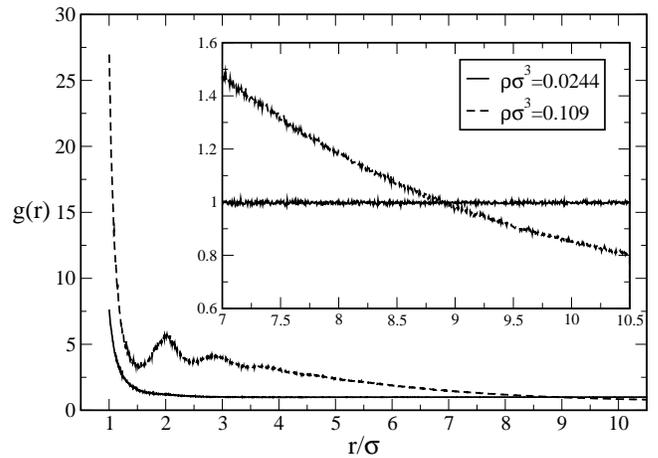}
\caption{The measured form of the radial distribution function $g(r)$ in
both the vapor phase and the spherical cluster phase that coexist at
the low density transition for $\epsilon^{-1}=0.41$. The inset shows an enlargement of the
behavior at large $r$.}
\label{fig:lowertrans_gr}
\end{figure}

The visual identification of clusters at the low density transition is
corroborated by measurements of the radial distribution function $g(r)$
of the coexisting phases. Fig.~\ref{fig:lowertrans_gr} displays the
measured forms of $g(r)$ for the vapor and the coexisting spherical
cluster phase at \ie$=0.41$. One notes that while the vapor phase is,
to a great extent, structureless (i.e.\ $g(r)$ effectively reaches its
asymptotic value of unity for $r \gtrsim 3 \sigma$), the cluster phase
exhibits unusual features, namely a very pronounced enhancement in the
value of $g(r)$ extending over a considerable length scale that is
indicative of the cluster radius. Furthermore, as the inset shows, $g(r
= 10.5 \sigma)\simeq0.8$ --significantly less than unity. This latter
feature heralds the onset of large-length-scale oscillations in $g(r)$
which have their origin in the inter-cluster correlations of the
cluster fluid phase \cite{ARCHER19}. Further evidence for this assertion
derives from our integral equation calculations to be presented in
Sec.~\ref{sec:ieqn}.

\begin{figure}[h]
\includegraphics[width=0.98\columnwidth,clip=true]{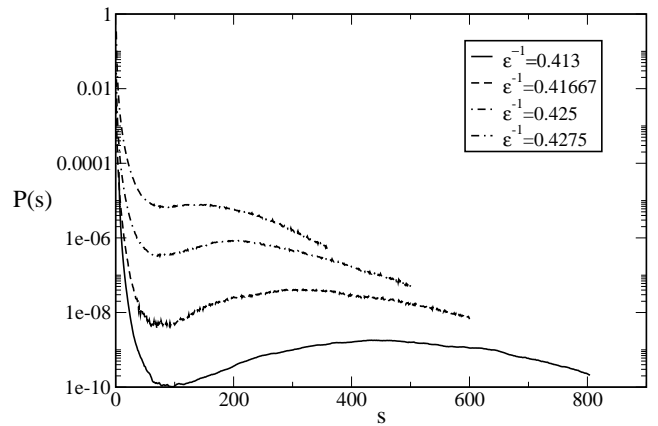}
\caption{Cluster mass distribution $p(s)$ (shown on a log scale) for a
selection of points along the vapor-cluster transition. The curves have
been shifted vertically to aid their distinguishability.}
\label{fig:clusterdist}
\end{figure}

The question naturally arises as to how the character of the spherical
cluster phase alters as one tracks the transition line by varying \ie.
To answer it,  in part at least, we have implemented a
cluster identification algorithm \cite{Hoshen76} and used it to obtain
the form of the cluster mass distribution $P(s)$ at the vapor-cluster
transition, for a number of coexistence state points. The results
(Fig.~\ref{fig:clusterdist}) show, in each instance, a double peaked
distribution. The peak at low mass occurs at $s=1$, indicating that
little or no clustering occurs in the vapor.  The broad peak occurring
at high mass indicates that the liquidlike clusters contain several
hundred particle. The latter peak shifts strongly to greater masses
with decreasing \ie, showing that the typical radius of the clusters
grows accordingly; this feature has important implications for
finite-size effects, as we discuss below. 

Finally in this section, we record that we have not attempted to
characterize in any detail the properties of the inhomogeneous
vapor-bubble phase occurring at the high density transition, due to the
prohibitively large computational effort that this would entail.

\subsection{Further inhomogeneous states}
\label{sec:inhom}

\begin{figure}[h]
\includegraphics[width=0.57\columnwidth,clip=true]{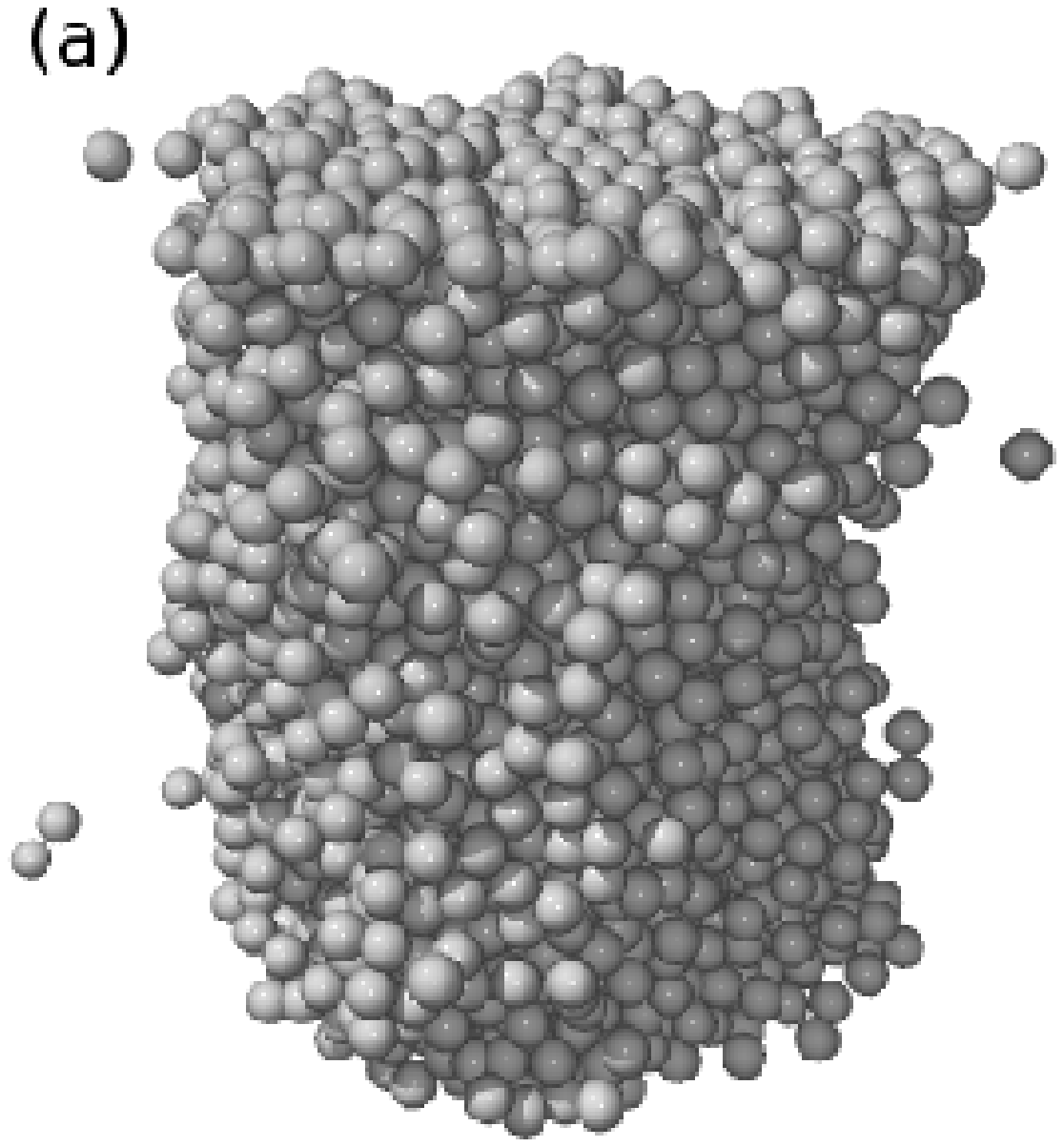}
\includegraphics[width=0.6\columnwidth,clip=true]{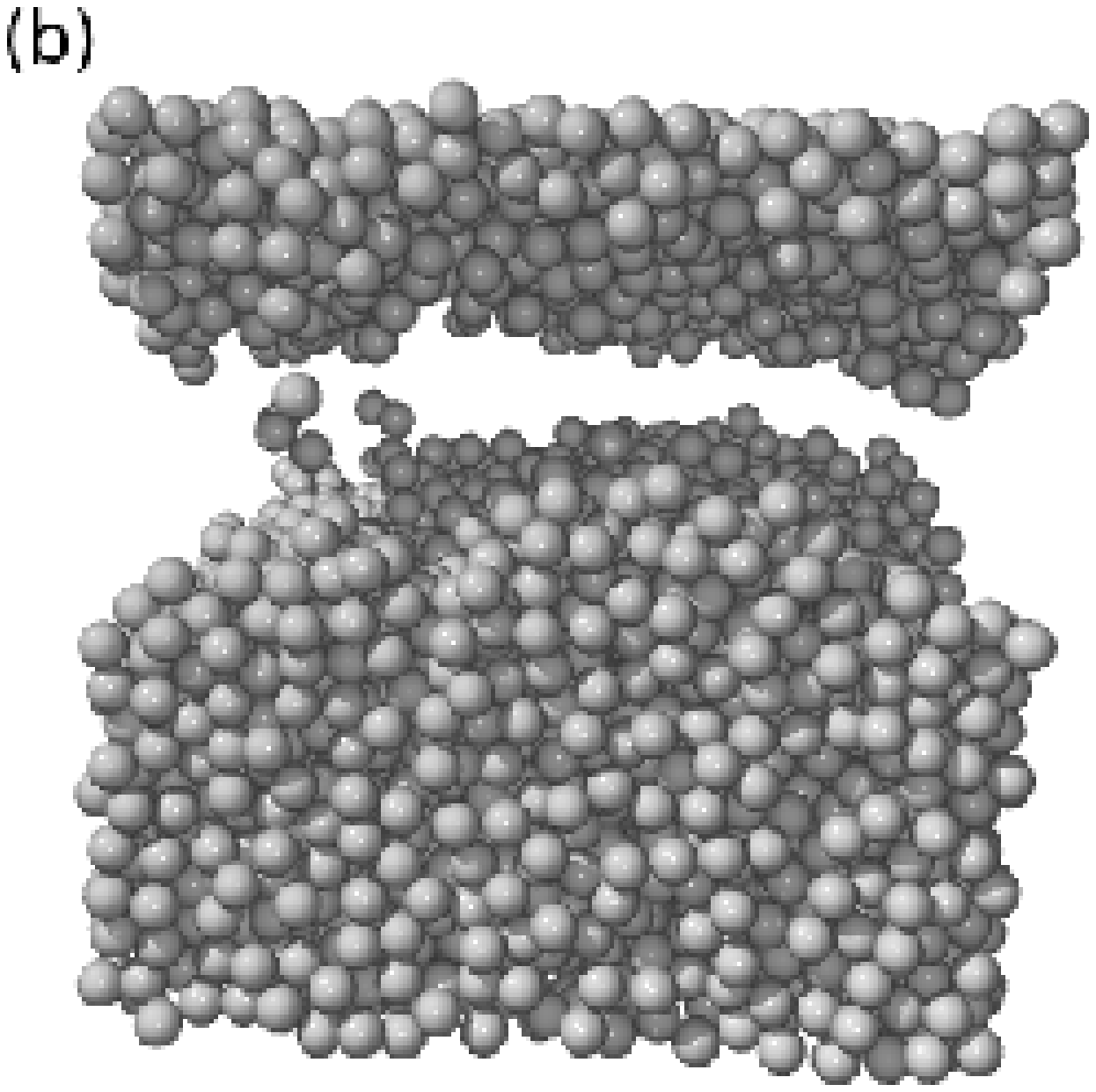}
\includegraphics[width=0.6\columnwidth,clip=true]{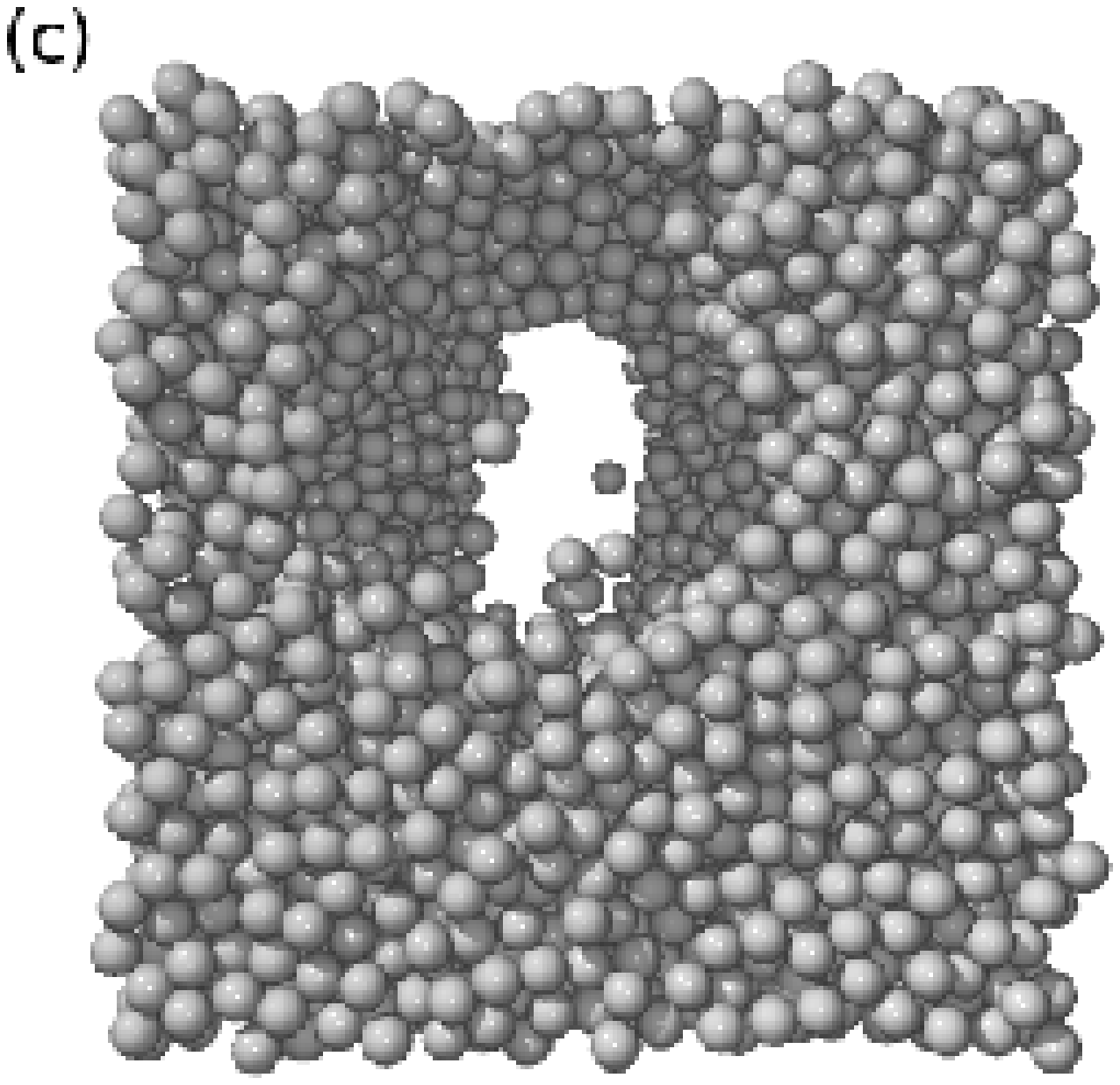}
\caption{Snapshots of the typical sequence of finite system
size-limited structures that were observed for \ie$=0.41$ when
traversing the region of densities intermediate between the density of
the spherical liquidlike cluster phase (which coexists with the vapor cf.\ Fig.~\ref{fig:coexist}(a)), and
the density of the spherical void phase (which coexists with the
liquid cf.\ Fig.~\ref{fig:coexist}(b)). {\bf (a)} A cylinder {\bf (b)} A slab {\bf (c)} A cylindrical
void. See text for more details.}
\label{fig:inhomo}
\end{figure}

We consider next the structures that form as one traverses the region
of inhomogeneous states separating the low and high density
transitions. On increasing $\mu$ at constant \ie$=0.41$ from its
coexistence value at the low density (vapor-cluster) transition,
visual inspection of typical configurations (Fig.~\ref{fig:inhomo}) shows
that the spherical clusters are replaced firstly by cylindrical
clusters (which span the periodic system in one direction), and then,
at still higher densities ($\bar\rho\sigma^3\simeq 0.2$), by slablike
structures (which span in two directions). In our simulations, one
structure appeared to evolve smoothly into the next, and we could
discern no clear signature of discontinuities in the gradient of the
$\bar\rho(\mu)$ in this range of densities (cf.
Fig.~\ref{fig:isotherms}(b)). Hence on this basis, there is no evidence
for the existence of first order phase transitions between spheres and
cylinders or cylinders and slabs. However, given that the
characteristic length scales of these structures is comparable with our
system size, we cannot rule out that sharp transitions could become
apparent for much larger system volumes.

Moving to higher densities ($\bar\rho\sigma^3\gtrsim 0.3$), two sharp kinks are
visible in the $\bar\rho(\mu)$ curve for \ie$=0.41$  (see
Fig.~\ref{fig:isotherms}(b)). The first kink, appearing at a density of
$\bar\rho\sigma^3\sim0.37$, occurs in a region in which the configurations are
slab-like, and appears to relate to a single cuboidal slab  being
replaced by two parallel slabs. Accommodating an additional slab within
the simulation box boosts the number of particles in each that are
within range of the long ranged repulsions from particles occupying the
other slab. Consequently it becomes energetically less favorable for
the slabs to thicken as $\mu$ is increased further, leading to a
decrease in the compressibility (and thus too, the gradient of
$\bar\rho(\mu)$). Accordingly, the kink at $\bar\rho\sigma^3\sim0.37$ is
something of a finite-size artifact. Visualization of configurations
either side of the second intermediate density kink (occurring at
$\bar\rho\sigma^3\simeq 0.47$) reveals that this is associated with a
transition from the parallel slabs of liquid just described, to a
liquid containing large cylindrically shaped voids, or ``bubbles'' of vapor which
span the system in one direction (Fig.~\ref{fig:inhomo}(c)). A 
jump in the density at $\bar\rho\sigma^3\simeq 0.7$ corresponds
to cylindrical voids being replaced by spherical voids.

Turning now to the data for \ie$=0.4$, shown in
Fig.~\ref{fig:isotherms}(b), the kinks that form the plateau in
$\bar\rho(\mu)$ at moderate densities have an origin distinct from that
just described for the case \ie$=0.41$. The configurations either side
of the first kink contain a single liquidlike slab. The plateau seems
to develop because --as the density grows-- the two surfaces of this
single slab are forced together via the periodic boundaries until, at
some point, the particles near one surface come within the range of the
repulsive part of the potential of the particles at the other surface.
Thus again, this feature would appear to be a finite-size artifact. The
second kink marks an abrupt condensation of the slab into the
homogeneous liquid phase. 

The character of the inhomogeneous structures that occupy the region
separating the pure phases helps to explain the dramatic increase in
the compressibility (and hence the gradient of $\bar\rho(\mu)$) on
reduction of \ie. This increase is (we believe) largely attributable to
the increasing prevalence of slab like structures as \ie\ is reduced.
Since the size of the spherical clusters that coexist with the pure
phases grows with decreasing \ie, they eventually approach that of the
system size, at which point spheres or cylinders are no longer stable
and are replaced by slabs. Since it costs little free energy to move a
slab interface (except when the two interfaces interact via the periodic
boundaries), the compressibility grows very large. We further note that
for values of \ie$\lesssim 0.39$,  i.e.\ below that of the triple
point, one should expect such behavior since at densities intermediate
between those of the stable coexisting vapor and liquid phases,
slab-like configurations dominate \cite{ERRINGTON03}.

Summing up the findings of this subsection, we have presented evidence
for a variety of inhomogeneous structures: spherical and cylindrical
liquidlike clusters, single and multiple liquidlike slabs, cylindrical
and spherical bubbles. However, the precise sequence of structures that
occurs when traversing the inhomogeneous region at fixed
$\epsilon^{-1}$ depends both on the value of $\epsilon^{-1}$ itself, and on the
system size. This is perhaps not altogether surprising, given that the
typical length scale of the inhomogeneous structures depends on
$\epsilon^{-1}$ (cf. Fig.~\ref{fig:clusterdist}), and that for
sufficiently small \ie, this length scale can exceed the linear extent
of the simulation box. Additionally, it should be mentioned that
protracted relaxation times were encountered in the inhomogeneous
region of the phase diagram. These arise because a local (i.e.\ single
particle) update procedure requires many iterations in order to
decorrelate a system having an inherently large length scale.
Accordingly, great computational expenditure is necessary to ensure
that the system attains the equilibrium structure for a given state
point. In fact, we observed a certain amount of irreproducibility
regarding the precise form of $\bar\rho(\mu)$ obtained on traversing
the inhomogeneous region from vapor to liquid compared to the reverse
path. The source of this irreproducibility is presumably traceable, in
part at least, to the extended relaxation times.

\section{Mean field and integral equation theoretical studies}
\label{sec:mf}
\subsection{Mean-field  theory of the fluid structure and thermodynamics}

\label{sec:rpa}

In Ref.\ \cite{ARCHER19} the authors developed a mean-field DFT theory
within the random phase approximation (RPA), for systems interacting
via potentials of the form in Eq.\ (\ref{eq:pair_pot}). We will not
describe in detail the theory here -- instead referring interested readers to
Ref.\ \cite{ARCHER19}. The key idea behind this approach (and indeed most
other mean-field approaches) is to split the pair potential into two
contributions: a reference part, $v_r(r)$ and the remainder or
`perturbation', $v_p(r)$, i.e.\ $v(r)=v_r(r)+v_p(r)$. For the present
system an obvious choice for the reference part is the hard sphere
potential, $v_r(r)=v_{hs}(r)$, where

\begin{equation}
v_{hs}(r) = 
\begin{cases}
\infty \hspace{10mm} r \leq \sigma \\
0 \hspace{12mm} r > \sigma,
\end{cases}
\label{eq:v_hs}
\end{equation}
and therefore:
\begin{equation}
\beta v_p(r) = 
\begin{cases}
-\epsilon+A \hspace{6mm} r \leq \sigma \\
\beta v(r) \hspace{9mm} r > \sigma.
\end{cases}
\label{eq:v_p}
\end{equation}
Note that there is no unique choice for $v_p(r)$ in the range $r \leq
\sigma$, inside the hard core. The choice made in Eq.~(\ref{eq:v_p}) is
the same as that of Ref.\ \cite{ARCHER19}. 

The Helmholtz free energy of the system is approximated as follows
\cite{ARCHER19}:

\begin{equation}
F[\rho] \simeq F_{hs}^{Ros}[\rho]+\frac{1}{2} \int \dr \int \dr' \rho(\rr) \rho(\rr') v_p(|\rr-\rr'|),
\end{equation}
where $\rho(\rr)$ is the one body density profile of the fluid and
$F_{hs}^{Ros}[\rho]$ is the Rosenfeld approximation for the Helmholtz
free energy of a hard sphere fluid \cite{RosenfeldPRL1989,
RosenfeldLevesqueWeisJCP1990, RosenfeldJCP1990}. For the bulk fluid this
amounts to the following (RPA) approximation for the pair direct
correlation function \cite{HM,ARCHER19}:

\begin{equation}
c(r;\rho)=c_{hs}^{PY}(r;\rho)-\beta v_p(r),
\label{eq:c_RPA}
\end{equation}
where $c_{hs}^{PY}(r;\rho)$ is the Percus-Yevick approximation for the hard-sphere direct pair correlation function for a bulk fluid of density $\rho$ \cite{HM,ARCHER19}.

\begin{figure}
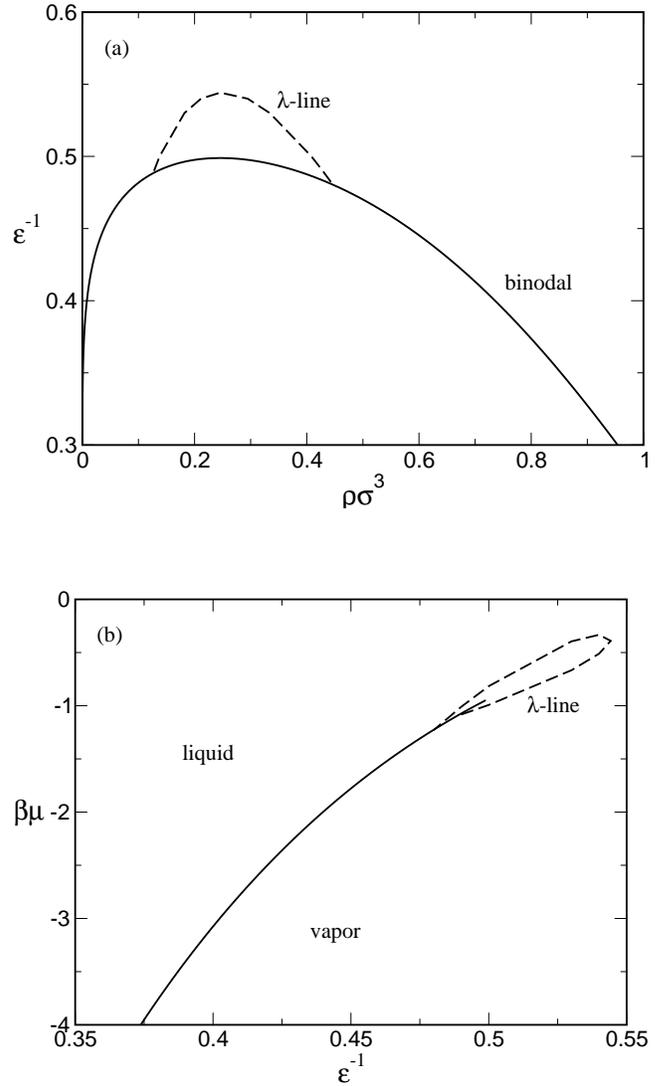

\includegraphics[width=0.98\columnwidth]{fig11a.eps}

\vspace{1cm}

\includegraphics[width=0.98\columnwidth]{fig11b.eps}
\caption{Phase diagram for $z_1=2$, $z_2=1$ and $A=0.55$, obtained from
the RPA DFT theory of Ref.\ \cite{ARCHER19}. The solid line is the
binodal and the dashed line is the $\lambda$-line. {\bf (a)} Plotted in the
density $\rho$ versus $\epsilon^{-1}$ plane, {\bf (b)} plotted in the
chemical potential $\mu$ versus $\epsilon^{-1}$ plane.}
\label{fig:RPA_DFT_result}
\end{figure}

For the potential parameters of concern in the present work ($z_1=2$,
$z_2=1$ and $A=0.55$), the mean-field theory of Ref.\ \cite{ARCHER19}
yields the phase diagram displayed in Fig.\ \ref{fig:RPA_DFT_result}.
From this figure one sees that the theory predicts a $\lambda$-line
enclosing the region of the phase diagram containing the liquid-vapor
critical point. The $\lambda$-line is defined as the locus of points in
the phase diagram for which the static structure factor $S(k)$ diverges
at a particular wave number $k_c \neq 0$ \cite{ARCHER19}. We take the
$\lambda$-line to indicate that the theory predicts a phase transition
to a periodically modulated inhomogeneous phase. This phase preempts
the standard liquid vapor behavior that normally occurs in the
subcritical region of a simple fluid. We note, however, that the
topology of the $\lambda$-line and hence the region of inhomogeneous
phase(s), is different from  that obtained from the MC simulations --
compare Figs.\ \ref{fig:coexcurve} and \ref{fig:RPA_DFT_result}(b) --
the $\lambda$-line suggests a closed loop around the inhomogeneous
phase(s). This seems to be inconsistent with the simulation results of
Sec.~\ref{sec:sim}, because the fluid of spherical clusters that we
have identified is not periodically modulated. Thus while the RPA
provides a reasonable approximation for the fluid structure and
thermodynamics in regions of the phase diagram away from the
$\lambda$-line \cite{ARCHER19}, it would appear to be unreliable in its
direct vicinity.

\subsection{Integral equation theories for the bulk fluid structure and thermodynamics}

\label{sec:ieqn}

The structure of a fluid may be characterized by the radial
distribution function $g(r)=1+h(r)$, where $h(r)$ is the total
correlation function, defined as the deviation of $g(r)$ from the ideal
vapor result \cite{HM}. The radial distribution function is important
for two main reasons: Firstly, it yields --via a Fourier transform--the static structure factor:

\begin{equation}
S(k)=1+\rho \int \dr \, (g(r)-1) \exp(i \kk \cdot \rr),
\label{eq:S_of_k}
\end{equation}
a quantity which, in principle, can be obtained in a scattering
experiment. Secondly, $g(r)$ can be used to calculate thermodynamic
quantities such as the internal energy and the pressure. For example,
the pressure $P$ can be obtained via the the virial equation \cite{HM}:

\begin{equation}
P=\rho k_B T - \frac{\rho^2}{6} \int \dr \, g(r) r \frac{{\rm d} v(r)}{{\rm d} r},
\label{eq:virial}
\end{equation}
while, the Helmholtz free energy may be obtained by thermodynamic
integration. In practice, however, since $g(r)$ is only approximately
known, the value obtained for the free energy is dependent upon the
path of integration \cite{CaccamoPhysRep1996}.

A key equation used to calculate $g(r)$ [or, equivalently $h(r)$] is the Ornstein-Zernike (OZ) equation \cite{HM}:
\begin{equation}
h(r)=c(r)+\rho \int \dr' c(|\rr-\rr'|)h(r'),
\label{eq:OZ_eq}
\end{equation}
where $c(r)$ is the pair direct correlation function. To solve the OZ equation it must be supplemented by a closure relation. The exact closure reads \cite{HM}
\begin{equation}
c(r)=-\beta v(r)+h(r)-\ln(1+h(r))-b(r),
\label{eq:exact_closure}
\end{equation}
where $-b(r)$ is the (unknown) bridge function. A simple approximation is to set $b(r)=0$. This is the HNC
closure \cite{HM}:
\begin{equation}
c_{HNC}(r)=-\beta v(r)+h(r)-\ln(1+h(r)).
\label{eq:HNC_closure}
\end{equation}
We solve Eqs.\ (\ref{eq:OZ_eq}) and (\ref{eq:HNC_closure}) to obtain
our approximation for $g(r)$. Another approximation we consider is the
PY closure \cite{HM}:

\begin{equation}
c_{PY}(r)=(1-\exp[\beta v(r)])(1+h(r)).
\label{eq:PY_closure}
\end{equation}
which is known to provide a good approximation for the structure and
thermodynamics of a hard sphere fluid at low and moderate
densities.

\begin{figure}
\includegraphics[width=0.98\columnwidth,clip=true]{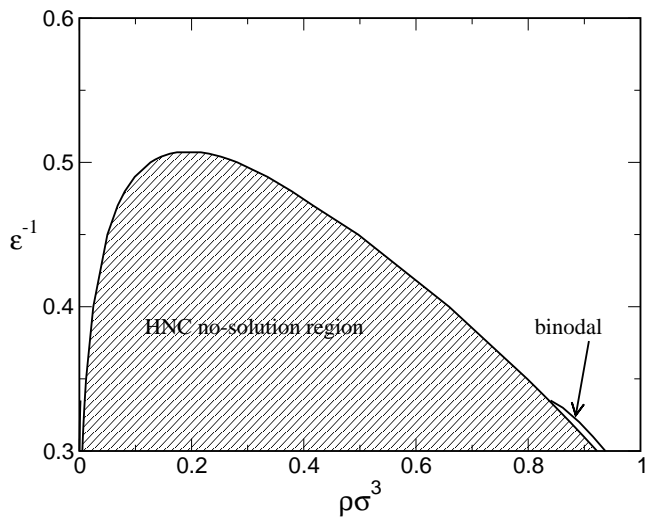}
\caption{The phase diagram in the \ie\ versus $\rho$ plane obtained
from the HNC integral equation theory for $z_1=2$, $z_2=1$ and
$A=0.55$. The shaded region denotes state points for which no solution
could be found. Outside this region, at low and moderate densities,
the HNC theory provides a good approximation for $g(r)$, when compared
with MC data. We find that at low values of $\epsilon^{-1}$ the HNC
predicts liquid-vapor coexistence.}

\label{fig:HNC}
\end{figure}

Within the HNC approximation, the chemical potential may be obtained
from the following expression:

\begin{eqnarray}
\beta\mu=\ln(\rho\Lambda^3)+\rho\int \dr \left\{\frac{h(r)}{2}[h(r)-c(r)]
-c(r)\right\},
\label{eq:chempot}
\end{eqnarray}
where $\Lambda$ is the (irrelevant) thermal de Broglie wavelength.
Coexistence between two phases occurs if they have equal temperature,
equal pressure and equal chemical potential. We employed Eqs.\
(\ref{eq:virial}) and (\ref{eq:chempot}) to calculate the fluid
pressure and chemical potential to find the coexisting densities within
the HNC approximation. The results are displayed in Fig.\
\ref{fig:HNC}. Unfortunately, only a small portion of the coexistence
binodal (at low values of $\epsilon^{-1}$) could be determined since
for the HNC approximation there is a large region of the phase diagram
in which there is no solution to the OZ equation. This region encompases
state points where we would expect (on the basis of the simulation) 
to find inhomogeneous phase(s) -- see Fig.\
\ref{fig:HNC}. We note that it is not unknown for the HNC
approximation to exhibit regions of no solutions
\cite{HoyeetalMolP1993, BelloniJCP1993, SchlijperetalJCP1992,
FerreiraetalJCP1994}. Indeed, one can obtain the HNC closure
(\ref{eq:HNC_closure}) as the Euler-Lagrange equation from an
associated free energy functional and it can be shown that the lack of
HNC solutions is due to the loss of convexity in this free energy
functional \cite{SchlijperetalJCP1992, FerreiraetalJCP1994}.

We have also applied a modified HNC (MHNC) closure in which we invoke a
simple approximation for the bridge function in Eq.\
(\ref{eq:exact_closure}), chosen to be the PY expression for the
hard-sphere fluid bridge function at the same density $\rho$. We found
that this MHNC approximation failed to converge for the same state
points as the HNC approximation. We did not attempt to implement any
other more sophisticated MHNC approximations \cite{CaccamoPhysRep1996}.

\begin{figure}
\includegraphics[width=0.98\columnwidth,clip=true]{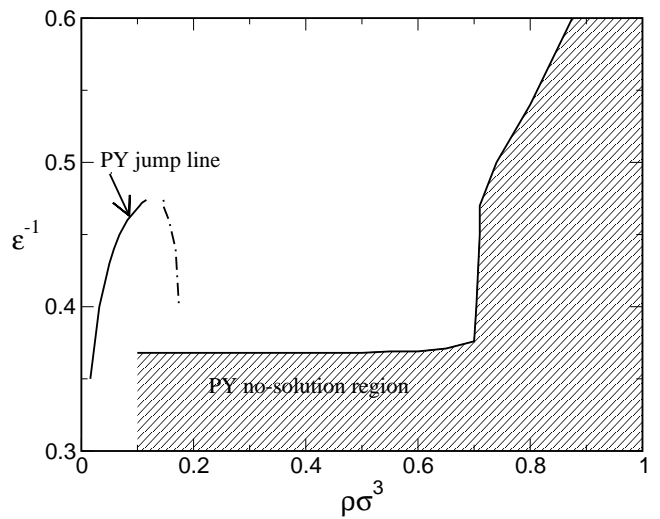}
\caption{The phase diagram in the \ie\ versus $\rho$ plane as obtained by 
the PY integral equation theory. The shaded region denotes state points 
for which no solution could be found. The solid line indicates
the line in the phase diagram at which there is a discontinuous change
in $g(r)$ -- compare with Figs.\ \ref{fig:g_of_r_Teq0_44}
and \ref{fig:g_of_r_Teq0_41}. This line is close the vapor-cluster
phase transition in our MC simulations. The dot-dashed line indicates
the locus of state point points at which the pressure [obtained via Eq.\
(\ref{eq:virial})] equals that in the vapor phase.}
\label{fig:PY}
\end{figure}

\begin{figure}
\includegraphics[width=1.\columnwidth,clip=true]{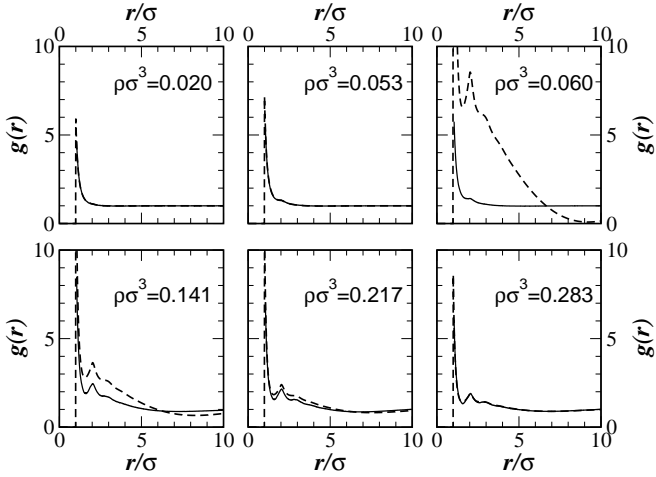}
\caption{The radial distribution function, $g(r)$, obtained from
simulation (solid lines) and the PY theory (dashed lines) for 
$\epsilon^{-1}=0.44$ at the particular selection of densities
indicated. Note that at $\rho \sigma^3=0.058$, the PY theory predicts a 
dramatic jump discontinuity in $g(r)$. An increase in the magnitude of
the first and second maxima also occurs in the MC simulation results at
a slightly higher density, but it is much less pronounced than that
predicted by the PY theory and does not appear to be discontinuous
for this value of $\epsilon^{-1}$.}
\label{fig:g_of_r_Teq0_44}
\end{figure}

\begin{figure}
\includegraphics[width=1.\columnwidth,clip=true]{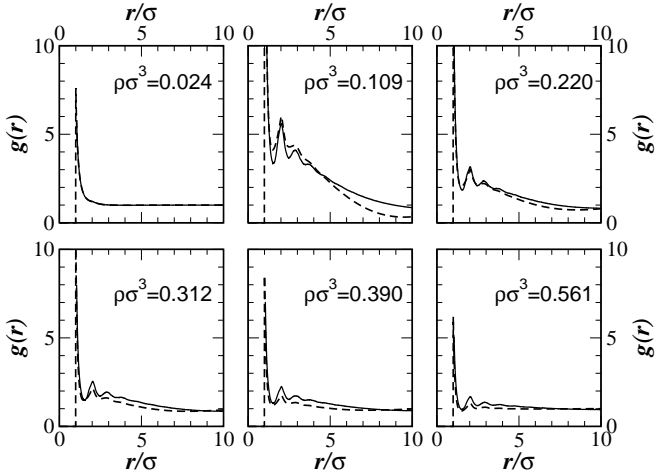}
\caption{The radial distribution function, $g(r)$, obtained from
simulation (solid lines) and the PY theory (dashed lines) for 
$\epsilon^{-1}=0.41$ at the particular selection of densities
indicated. Note that in both the simulations and theory, a
jump discontinuity occurs in $g(r)$ between $\rho\sigma^3=0.024$ and
$\rho\sigma^3=0.109$. Within the PY theory, this jump occurs at $\rho
\sigma^3=0.04$.}

\label{fig:g_of_r_Teq0_41} 
\end{figure}

Applying the PY closure (\ref{eq:PY_closure}) to the OZ equation
(\ref{eq:OZ_eq}) gives some (perhaps) surprising results: As with the
HNC, there are regions of the phase diagram where we were unable to
obtain a solution. However, their location is different to that
found for the HNC -- compare Figs.\ \ref{fig:HNC} and \ref{fig:PY}. The
theory also seems to succeed in capturing some key aspects of the cluster
transition. Specifically, if for some constant $\epsilon^{-1}\lesssim 0.48$, one
calculates the PY approximation for $g(r)$ along a path of increasing
density, starting from very low densities, one finds at some $\rho$
that a discontinuous jump occurs in the solution for $g(r)$-- see for
example the results in Figs.\ \ref{fig:g_of_r_Teq0_44} and
\ref{fig:g_of_r_Teq0_41}. The locus of points in the phase diagram at
which $g(r)$ `jumps' in this way is displayed in Fig.\ \ref{fig:PY}. Note also,
that if one traverses the reverse path, one can follow the high density
branch of solutions to the PY displaying the clustering to very low
densities. The low density solutions on this branch are particularly
striking. The location in the phase diagram of the `jump' line is close
to where the transition to a cluster phase occurs, as determined from
our MC simulations.

\begin{figure}
\includegraphics[width=1.\columnwidth,clip=true]{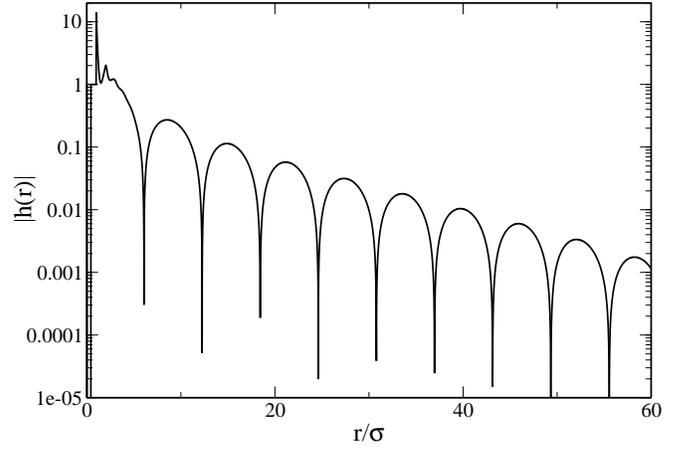}
\caption{The modulus of the correlation function $h(r)$ obtained from
the PY theory for $\epsilon^{-1}=0.41$ and $\rho \sigma^3=0.220$.}
\label{fig:mod_h} 
\end{figure}

In Fig.\ \ref{fig:mod_h} we display $|h(r)|$ obtained from the PY
theory for the state point $\epsilon^{-1}=0.41$ and $\rho
\sigma^3=0.220$ (see also Fig.\ \ref{fig:g_of_r_Teq0_41}). The
logarithmic vertical axis allows inspection of the long wavelength
oscillatory decay of this correlation function, arising from
inter-cluster correlations. A precursor to this oscillatory behavior
is visible in our simulation results (Fig.~\ref{fig:lowertrans_gr}).
We note that the amplitude of the oscillations decays rather slowly
indicating very long ranged correlations in the cluster phase. 

We have attempted to calculate the binodal of the vapor-cluster
transition from PY theory. Initially we sought to perform thermodynamic
integration of the pressure (as obtained from the virial
\ref{eq:virial}) to acquire the Helmholtz free energy
\cite{CaccamoPhysRep1996} and thence the coexisting state points.
However, we were unsuccessful in this endeavor due to insufficient
self consistency within the theory. As an alternative, more
approximate, approach we have done the following: i) We assumed that
the `jump' line obtained from the PY theory (see Fig.\ \ref{fig:PY}) is
roughly the coexistence curve for the vapor phase -- our MC simulation
data supports this assumption. ii) To find the coexisting cluster phase
density, we sought the cluster phase state point having the same value
of $\epsilon^{-1}$ and pressure [obtained via Eq.\ (\ref{eq:virial})]
as the vapor phase at densities just below the jump line. This approach
yields the dot-dashed line in Fig.\ \ref{fig:PY}. Similar results can
be obtained from an alternative approach, which is to plot the pressure
obtained from the virial equation (\ref{eq:virial}) as a function of
density for fixed $\epsilon^{-1}<0.477$; this produces a curve
exhibiting a van der Waals loop (as well as a discontinuity at the jump
line). Ignoring the discontinuity, one can perform the equal areas
construction in order to obtain coexisting state points.

\begin{figure}
\includegraphics[width=0.98\columnwidth,clip=true]{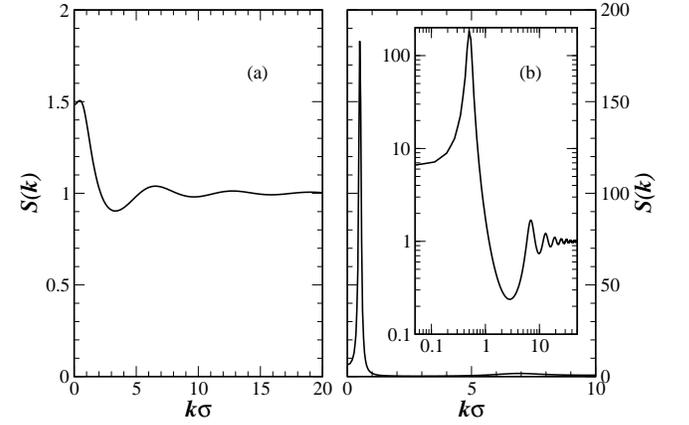}
\caption{The static structure factor, $S(k)$, obtained for the PY
theory for the state points $A=0.55$, $\epsilon^{-1}=0.41$ and for densities
{\bf (a)} $\rho \sigma^3=0.024$, {\bf (b)} $\rho \sigma^3=0.220$. For
both these state points the agreement between the PY results and the MC
simulations for $g(r)$ is good -- see Fig.\ \ref{fig:g_of_r_Teq0_41}.
The inset to {\bf (b)}, shows $S(k)$ on a logarithmic scale.}
\label{fig:S_of_k}
\end{figure}

In Fig.\ \ref{fig:S_of_k} we display the static structure factor $S(k)$
obtained from the PY theory for two state points either side of the
vapor cluster-phase transition line. For these state points the PY
result for $g(r)$ agrees well with our MC simulation result -- see
Fig.\ \ref{fig:g_of_r_Teq0_41}. $S(k)$ displays a peak at small wave
vector $0<k_c \ll 2 \pi/\sigma$ for states both sides of the transition
line. This peak is known to be a signature of clustering. However, in
the cluster phase the peak height is several orders of magnitude
greater than in the vapor phase. 

\begin{figure}
\includegraphics[width=0.98\columnwidth,clip=true]{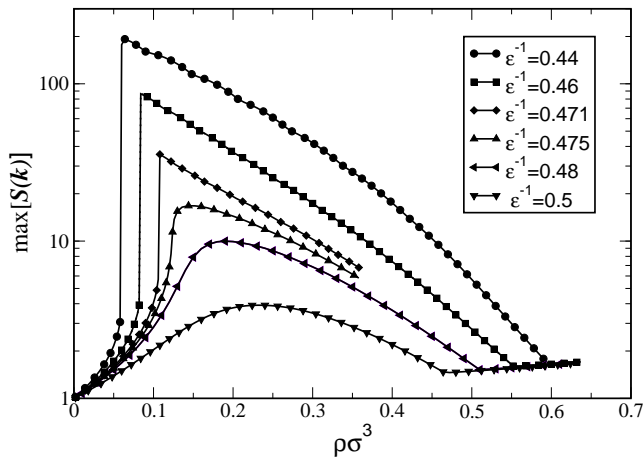}
\caption{The behavior of the maximal value of the static structure factor $S(k)$ as a
function of $\rho$, obtained within the PY theory for a number of different values of
$\epsilon^{-1}$, as described in the text.}
\label{fig:S_MAX}
\end{figure}

Fig.\ \ref{fig:S_MAX} shows the peak (maximal) value of $S(k)$ obtained
from the PY theory as a function of the fluid density $\rho$, for a
selection of values of $\epsilon^{-1}$. At low $\rho$, $S(k) \simeq 1
\: \forall k$. As the density is increased, the peak that grows the
fastest, is that at $k=k_c$, i.e.\ the curves in Fig.\ \ref{fig:S_MAX} represent the dependence of
$S(k_c)$ on $\rho$ in this regime. For values of $\epsilon^{-1}<0.48$,
a jump occurs in $S(k_c)$ at the density of the vapor-cluster
transition. Underlying this jump in the peak height is a wholesale
discontinuous change in the entire form of $S(k)$. As we have
previously shown \cite{ARCHER19}, the value of the structure factor at
$k=0$ is roughly proportional to its value at $k=k_c$, and
consequently, $S(k=0)$ also jumps at the transition point. Given,
however, that $S(0)=\rho k_BT \chi_T$, where $\chi_T$ is the isothermal
compressibility, this in turn implies that a jump discontinuity occurs
in $\chi_T$ at the transition density, in accordance with the
simulation results.

As the density is increased past the value at which $S(k)$ jumps as a
whole, the value of $S(k_c)$, ie. the peak height decreases.
Eventually, at some higher density (the precise value depending on the
particular value of $\epsilon^{-1}$), the maximal value of $S(k)$ no
longer occurs at $k=k_c \neq 0$, but instead at $k \simeq 2
\pi/\sigma$, the next maximum in $S(k)$. This next maximum originates
from the hard-sphere correlations in the fluid, and the crossover in
maximal value from one peak in $S(k)$ to the other, accounts for the
higher density discontinuity in the gradient of the curves displayed in
Fig.\ \ref{fig:S_MAX}. For example, for the case when
$\epsilon^{-1}=0.44$, this occurs at $\rho \sigma^3 \simeq 0.6$.

\begin{figure}
\includegraphics[width=0.98\columnwidth,clip=true]{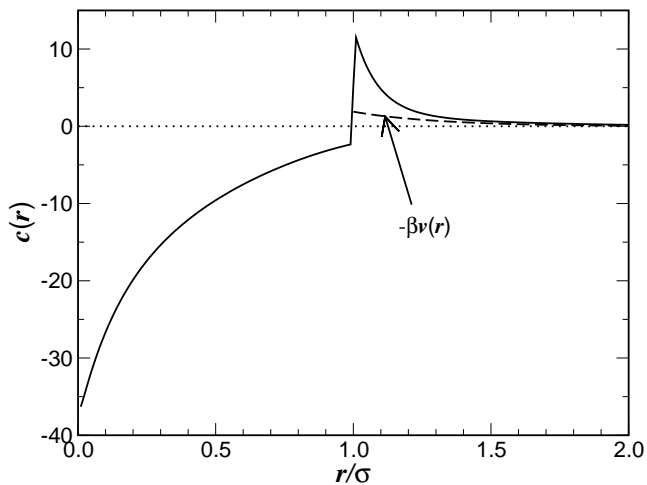}
\caption{The direct pair correlation function $c(r)$ obtained from the
PY theory for the parameters $A=0.55$ and $\epsilon^{-1}=0.41$ and
density $\rho \sigma^3=0.220$. The dashed line is a plot of $-\beta
v(r)$. For the corresponding $g(r)$, see Fig.\
\ref{fig:g_of_r_Teq0_41}, and for the corresponding $S(k)$, see Fig.\
\ref{fig:S_of_k}.}
\label{fig:c_of_r}
\end{figure}

In Fig.\ \ref{fig:c_of_r} we display the direct pair correlation
function for the state point $A=0.55$, $\epsilon^{-1}=0.41$ and density
$\rho\sigma^3=0.220$. For this state point the PY approximation for
$g(r)$ is in good agreement with the results from our MC simulations --
see Fig.\ \ref{fig:g_of_r_Teq0_41}. Note that for $1<r/\sigma<1.3$,
$c(r)\gg -\beta v(r)$. We believe it is for this reason that the MSA
closure to the OZ equation:

\begin{eqnarray}
h(r)=-1, \hspace{9mm} r< \sigma, \notag\\
c(r)=-\beta v(r), \hspace{3mm} r> \sigma,
\label{eq:MSA}
\end{eqnarray}
completely fails to describe the fluid structure in the region of the phase diagram near the
vapor-cluster phase transition -- the MSA approximation forces $c(r)$ to
be much smaller than it is in reality for $1<r/\sigma<1.3$. We believe that it is also for this reason that the RPA closure, Eq.\ (\ref{eq:c_RPA}),
is unreliable in the vicinity of the vapor-cluster transition.

\section{Discussion}
\label{sec:disc}

In summary, we have studied the phase behavior of a system of particles
which interact via a mermaid potential, as a function of the strength
of the short ranged attraction, for a fixed strength of the long ranged
repulsion. The effect of the repulsion is to substitute the
liquid-vapor critical point and a portion of the associated
liquid-vapor transition line, with two first order phase transitions:
one from the vapor to a fluid of spherical liquidlike clusters, and the
other from the liquid to a fluid of spherical voids. Each of these
phases is highly compressible compared to the respective homogeneous
phase with which it coexists. At low temperature, the two transition
lines intersect one another and the vapor-liquid transition line at a
triple point, while at high temperatures, they appear to terminate at
distinct critical points.  Although SCOZA, HRT and most of the standard
integral equation theories are unable to describe the new transitions,
somewhat surprisingly the Percus Yevick approximation does succeed in
capturing the vapor-cluster phase transition, as well as key aspects of
the structure of the cluster fluid.

\begin{figure}
\includegraphics[width=0.95\columnwidth,clip=true]{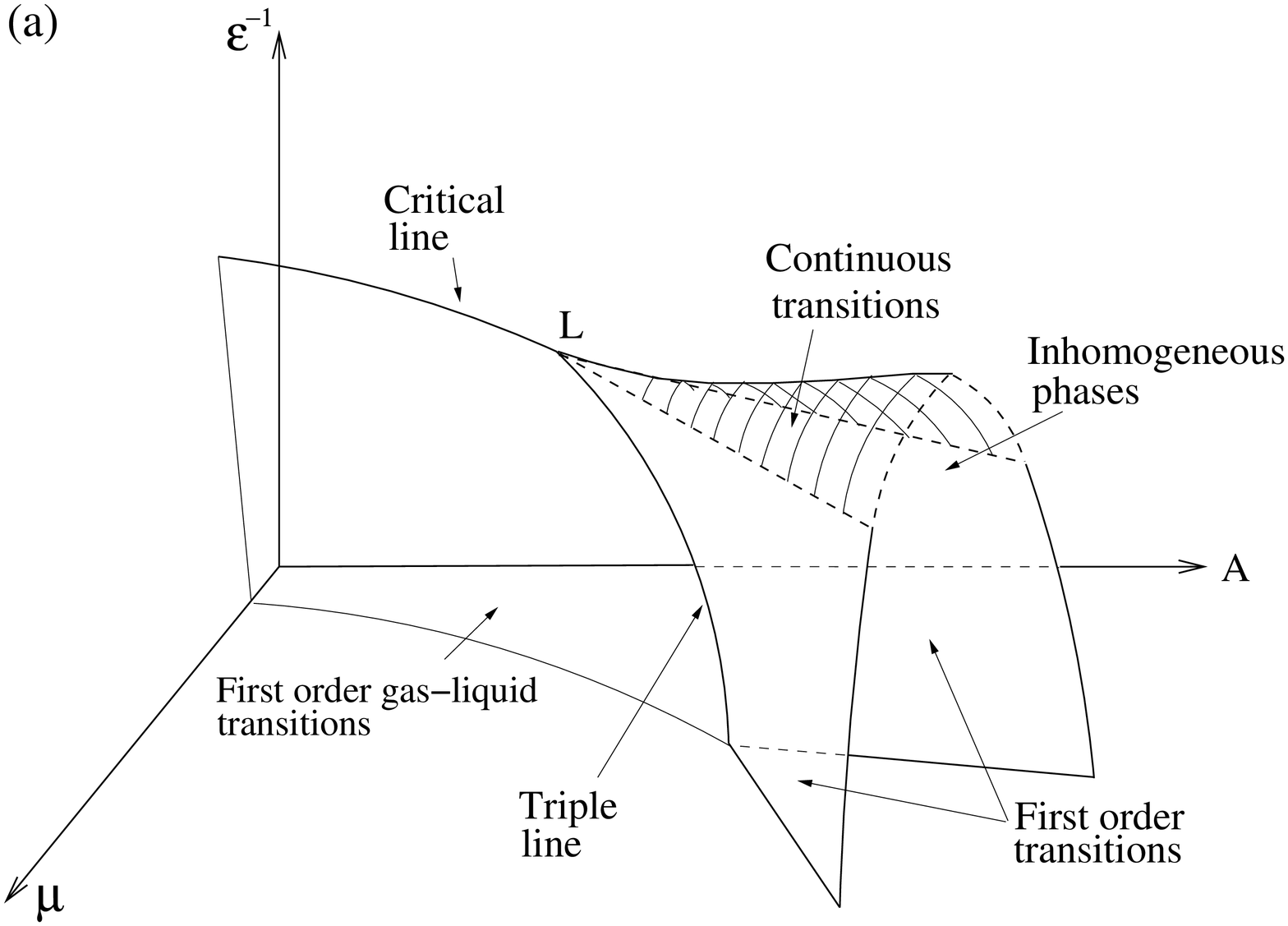}
\includegraphics[width=0.95\columnwidth,clip=true]{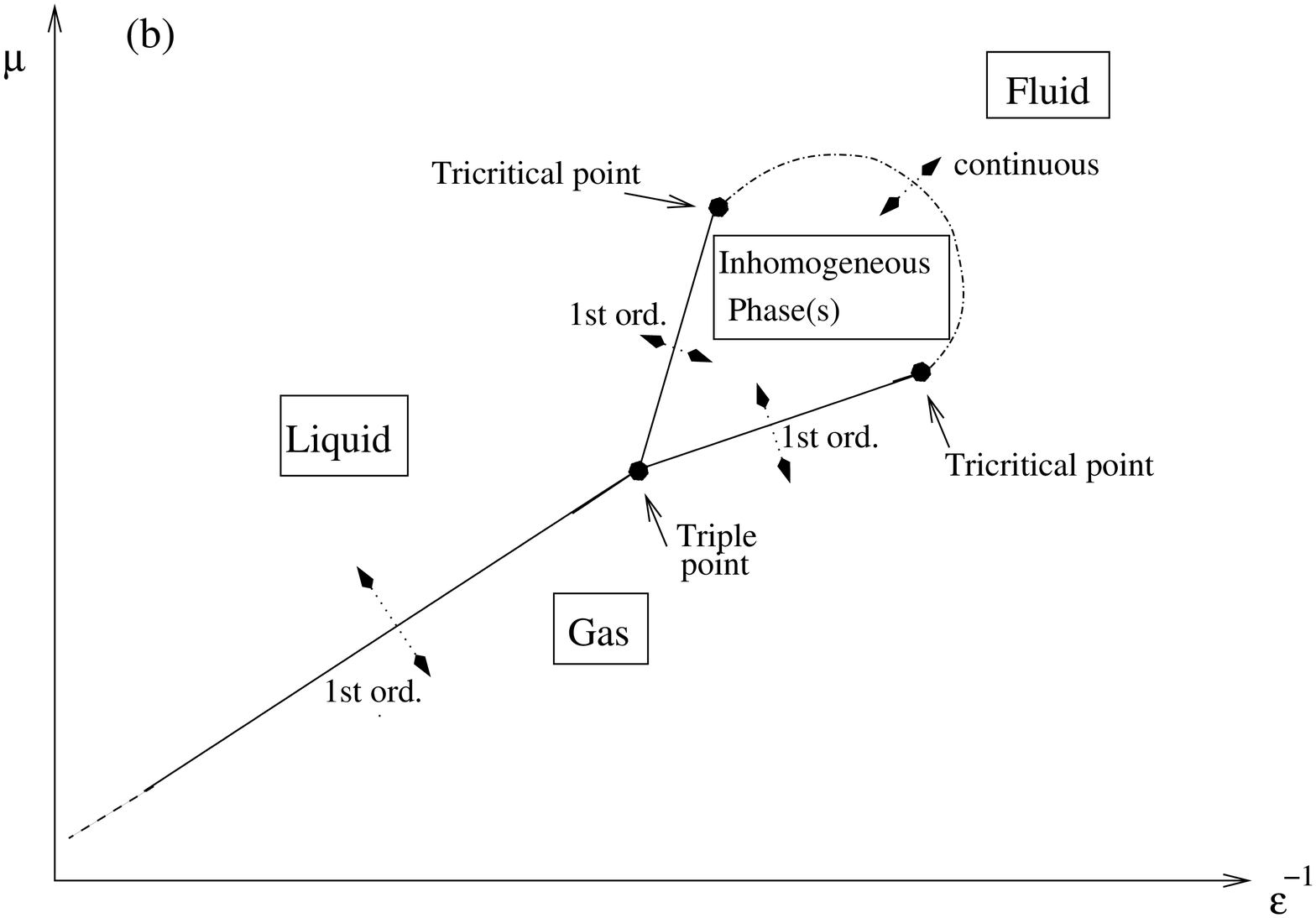}
\caption{{\bf (a)} A schematic representation of a possible form of the full
phase diagram for the model of Eq.~(\ref{eq:pair_pot}), adapted from
ref.~\cite{Barbosa93} and described in the text. {\bf (b)} A cut at constant
$A\gtrsim A_L$ through the phase diagram of {\bf (a)}.}
\label{fig:schematic}
\end{figure}

Owing to the high computational cost of the current study, we have
obtained the phase behaviour only for a single value of the strength of
the repulsion namely $A=0.55$. Consequently, it is difficult to comment
authoritatively on the topology of the phase diagram in the full space
of $\mu, \epsilon$ and $A$. Nevertheless, it is worthy of note that the
phase diagram of Fig.~\ref{fig:coexcurve} is largely consistent with a
cut at constant $A$ through a phase diagram originally obtained by
Barbosa in a mean field study of an Ising model having isotropic
competing interactions \cite{Barbosa93}. Fig.~\ref{fig:schematic}(a)
recasts Barbosa's phase diagram in terms appropriate for the present
fluid model. One sees from the figure that as $A$ is increased from
zero, the liquid-vapor critical point shifts to lower \ie\ in
accordance with SCOZA and HRT studies \cite{PinietalCPL2000} of mermaid
potentials. At some value of $A=A_L$ -- corresponding to a Lifshitz
point-- inhomogeneous phases start to occur and two sheets (or wings)
of first order phase transitions emerge from a triple line to delineate
a region of inhomogeneous states. This region is capped off at high
\ie\ by a surface of continuous transitions. A cut through the phase
diagram at constant $A\gtrsim A_L$ has the form shown in
Fig.~\ref{fig:schematic}(b), which is qualitatively very similar to our
Fig.~\ref{fig:coexcurve}, except that we have not, as yet, been able to
identify a line of continuous transitions.

A phase diagram of the form Fig.~\ref{fig:schematic}(a) might shed
light on the observation by Pini et al.\ of an anomalously large region
of high compressibility around the critical point
\cite{PinietalCPL2000}. If the strength of the repulsion is such that
$A\lesssim A_L$, then the system will exhibit a liquid-vapor critical
point, but will be located (in the full phase diagram) close to the
Lifshitz point. As we have seen in the present work, the inhomogeneous
states that form for $A>A_L$ are highly compressible compared to the
vapor or the liquid, and even though they aren't stable for $A\lesssim
A_L$, they would be expected to have an influence on the free energy
landscape. This could account for the anomalously large region of high
compressibility.

As described in Sec.~\ref{sec:inhom}, when traversing the inhomogeneous
region separating the low and high density transitions, a variety of
finite-size limited (spanning) structures were observed depending on
the value of $\epsilon$ and the system size. Evidence was found for
cylinders and slabs, as well as cylindrical voids. It is interesting
to note that a similar sequence of inhomogeneous structures has also
been reported in simulations of the sub-critical Lennard-Jones fluid
using periodic boundary conditions \cite{MACDOWELL06}. There the
observed structures occurred for values of the density intermediate
between the coexisting stable vapor and liquid phases, and were
metastable with respect to these phases. It therefore seems that
the effect of adding a long ranged repulsion to a purely attractive
fluid, is to stabilize these structures with respect to the homogeneous
phases. We further note that while the clusters that occur at the low
and high density transitions have a length scale smaller than our
system size (at least for larger \ie), and are therefore expected to
persist in the thermodynamic limit, our system sizes are too limited to
make definitive statements regarding the situation at moderate
densities. Here it seems more likely that modulated structures occur,
the wavelength of which exceeds our linear system sizes.

With regard to the prospects for extensions to the present work, we
were not able to satisfactorily address the question of the nature of
the putative critical points which terminate the vapor-cluster and
liquid-cluster transitions. To do so will the require the
identification of the appropriate order parameter for the transitions,
and a reformulation of finite-size scaling methodologies to take
account of the large cluster length scale \cite{archerfootnote4}. Once
this is achieved, the question as to whether a line of continuous
transitions really does link the (tri)critical points (cf.
Fig.~\ref{fig:schematic}(b)) could be tackled. As for future integral
equation studies, another closure approximation for the OZ equation, in
common employ, is the Rogers-Young (RY) closure
\cite{CaccamoPhysRep1996, RogersYoungPRA1984}. This interpolates
between the PY and HNC theories in such a way as to enforce consistency
between the virial and the compressibility routes to the
thermodynamics. Since the PY theory is able to describe the vapor
cluster-phase transition, in contrast to the HNC, it would be
interesting to see how the RY closure fares in this respect.

\acknowledgments

The authors have benefited from helpful discussions with R. Evans, D.
Pini and L. Reatto. AJA gratefully acknowledges support from RCUK.

\bibliography{nbwrefs}

\begin{thebibliography}{10}
\bibitem{archerfootnote4}
The computational problems that stem from the large typical cluster size at
  criticality might be ameliorated by choosing a different amplitude of the
  repulsion such as to reduce the characteristic length scale of the cluster
  instability.
\end{thebibliography}

\bibliographystyle{prsty}

\end{document}